\documentclass[journal,12pt,onecolumn,draftclsnofoot,]{IEEEtran}
\usepackage[colorlinks,urlcolor=blue,linkcolor=blue,citecolor=blue]{hyperref}
\usepackage{color,array}
\usepackage{graphicx}
\usepackage{amsmath}

\usepackage{mathtools}
\usepackage{bbm}
\usepackage{hyperref}
\usepackage{amssymb}
\usepackage{amsthm}
\usepackage{charter}
\usepackage{graphicx}
\usepackage{bm}

\DeclareMathOperator*{\argmaxA}{\arg\max} 
\DeclareMathOperator*{\argminA}{\arg\min}

\usepackage{algorithmic}
\usepackage{booktabs}

\usepackage{graphicx}
\usepackage{mathrsfs}
\usepackage[justification=centering]{caption}
\usepackage{breqn}

\newtheorem{theorem}{Theorem}

\newtheorem{definition}{Definition}

\usepackage{enumitem}

\begin{document}

%MRB: Change MD -> Auctions or add some other (non-auction) mechanisms / see comment in Section 1 last but one para
\title{Deep Learning Meets Mechanism Design: A Survey}

\author{V. Udaya Sankar, Vishisht Srihari Rao, Mayank Ratan Bhardwaj, and Y. Narahari
\thanks{The first author gratefully acknowledges the support provided by the  TARE (Teachers Associateship for Research Excellence) Grant of the Department of Science and Technology, Government of India.}
\thanks{\textbf{Corresponding Author}:V. Udaya Sankar is with the Department of Electronics and Communication Engineering, SRM University A.P., Guntur, Andhra Pradesh, India. (e-mail: udayasankar.v@srmap.edu.in)}
\thanks{Vishisht Srihari Rao is with Carnegie Mellon University, Pittsburgh, Pennsylvania, USA (email: vsrao@cs.cmu.edu)}
\thanks {Mayank Ratan Bhardwaj is with Plaksha University, Mohali, India (email: mayank.bhardwaj@plaksha.edu.in)}
\thanks {Y. Narahari is with the Department of Computer Science and Automation, Indian Institute of Science, Bangalore, India (Email: narahari@iisc.ac.in)}
}

% \markboth{}{Sankar, V.U., Rao, V.S., \& Narahari, Y.}{DL Meets MD}

\maketitle

\begin{abstract}
Mechanism design is a topic of much importance in economic theory and can be described as the reverse engineering of games, involving the design of rules for strategic agents in such a way that the resulting game achieves specific desirable properties in an equilibrium of the game. 
%Auctions and matching markets are prominent examples of mechanisms. 
Mechanism design has been extensively used in numerous engineering applications, such as the design of auctions, the design of matching markets, Internet advertising, network design, and social network analysis. Key desirable properties for mechanisms include incentive compatibility, individual rationality, budget balance, welfare maximization, revenue maximization (or cost minimization), and fairness in allocation. However, mechanism design theory asserts that only certain strict subsets of these properties can be achieved simultaneously by any given mechanism. In practice, however, mechanisms addressing real-world challenges often require satisfying combinations of properties that are theoretically infeasible. To address this gap, a recent approach uses deep learning to {\em learn\/} mechanisms that satisfy the required properties. This is achieved by minimizing a suitably defined loss function during the training of an appropriate deep learning network. In this paper, we describe  the core technical details of this deep learning-based approach to mechanism design and provide a comprehensive survey of the key methods which have emerged in the recent years. We further illustrate the effectiveness of this approach through three real-world engineering applications: (a) energy management in vehicular networks, (b) resource allocation in mobile networks, and (c) volume discount procurement auctions for agricultural inputs in digital agriculture.
\end{abstract}

%\begin{IEEEImpStatement}
%The impact statement should not exceeed 150 words. This section offers an example that is expanded to have only and just 150 words to demonstrate the point. Here is an example on how to write an appropriate impact statement: Chatbots are a popular technology in online interaction. They reduce the load on human support teams and offer continuous 24-7 support to customers. However, recent usability research has demonstrated that 30\% of customers are unhappy with current chatbots due to their poor conversational capabilities and inability to emotionally engage customers. The natural language algorithms we introduce in this paper overcame these limitations. With a significant increase in user satisfaction to 92\% after adopting our algorithms, the technology is ready to support users in a wide variety of applications including government front shops, automatic tellers, and the gaming industry. It could offer an alternative way of interaction for some physically disable users.
%\end{IEEEImpStatement}


\section{Introduction}
How do you realize social goals in a modern society consisting of self-interested individuals or agents?
This is a question facing all social planners and organisations all the time. If we solve this problem
satisfactorily, it has far-reaching implications for creating robust sociological institutions, as
well as, for solving numerous modern problems arising at the interface of computer science and
microeconomics. 
Mechanism Design is an important sub-field of economic theory that strives to answer the above question. Mechanism design can be thought of as the reverse engineering of a game to achieve  specific desired objectives. In mechanism design, based on the objectives to be achieved, the mechanism induces a game in a way that the desired, specific objectives are realized in an equilibrium of the game \cite{BORGERS15, NISAN07, NARAHARI14, NARAHARI09}. 
Auctions and matching market protocols are important and popular examples of mechanisms. In this paper, we focus on auctions; however, the results presented apply to more general mechanisms as well.

%MRB: Make first letter of second column capital? Remove 2nd part of caption in tables 1,2 for more space
\subsection{Motivation for the Survey}
The following are widely accepted as  the most desirable properties to be satisfied by an auction mechanism (these will be covered in more detail in the next section).  
incentive compatibility  (IC),
%\\Incentive compatibility ensures truthful bidding by the players and is a fundamental requirement of any auction mechanism. The most powerful version of IC is dominant  strategy incentive compatibility (DSIC), which means bidding true values is best irrespective of the bids of the other players.
%\item 
individual rationality (IR), budget balance, 
%\\Individual rationality ensures that the players obtain non-negative utility by participating in the auction. The most powerful version of IR is ex-post IR, which implies that, at the end of the auction, the utility to each participating player will be non-negative irrespective of the bids of the other players.
social welfare maximization (SWM), 
%\\This implies maximizing the cumulative utility of all the players in the auction.%sum of utilities of all the players in the auction. 
revenue maximization (or cost minimization),  fairness of allocation, etc.
%\\This means the expected total revenue for the seller (in the case of a forward auction) or the expected total cost of procurement for the buyer (in the case of a reverse auction) is optimized by the auction. Such auctions are also called optimal auctions.
%Commented out FAIR and BUS as (1)they are covered elsewhere, (2)even the other 4 are impossible, and (3)Not all papers have BUS
%\item {\bf Fairness (FAIR)}   \\Fairness implies that the winning bidders are chosen in a fair way. An index of fairness would be envy-freeness -- no bidder  can increase their utility by adopting another bidder's outcome. If envy-freeness is not achievable, the next best option is envy minimization.
%\item {\bf Business constraints (BUS)}   \\Satisfaction of business constraints refers to constraints such as having a minimum number of winning bidders (to avoid monopoly),  a maximum number of bidders (to minimize operational complexity), a maximum fraction of business to be awarded to any bidder, etc. 
%\end{enumerate}

%MRB:Mention in this para: Similarly, this DL approach can be applied to all mechanisms (after making necessary tweaks), although we have only taken examples of auctions. [OR] Change the paper's title from MD to Auctions
%VR: Change the below line as it is no longer 6 properties.
Satisfying all of the properties listed above simultaneously is clearly a tall order. Mechanism design theory \cite{NARAHARI14, MASCOLELL95, KRISHNA09} is replete with celebrated results which assert that these properties  cannot be simultaneously satisfied.  This has led to several methodologies to surmount this limitation. One of the recently emerged, novel  methodologies is an approach based on deep learning (DL). The  goal of the current survey  paper is to  discuss this recent approach, based on Deep Learning, for designing mechanisms  that achieve maximal subsets of these objectives, approximately,  with as little compromise (or regret) as possible. 
The  DL-based approach achieves this by   learning a mechanism that approximately satisfies the required set of properties by minimizing a suitably defined loss function. Additionally, the DL-based approach may reduce the computational complexity of deducing the mechanism. In this paper, we present key technical details of the DL-based approach for mechanism design and provide an overview of the key methods that have emerged in this topic. 

\subsection{Outline of the Paper}
The rest of this paper is organized as follows. Section \ref{sec:ClassicalResults} presents some essentials of mechanism design that are needed for the rest of the paper. We first define mechanisms and introduce auctions as representative examples of mechanisms. We next describe briefly the desirable properties that mechanisms are required to satisfy and state a few possibility and impossibility results. This provides the motivation for the main subject of this paper, namely, a deep learning based approach for mechanism design. In Section \ref{sec:DLMDExample}, we discuss an illustrative example of an online ad-auction to motivate the deep learning based approach.  Sections \ref{sec:revopt}-\ref{sec:BudgetBalanced} are devoted to four different groups of methods based on four different criteria, namely,  revenue maximization; welfare maximization; fairness of allocation; and budget balance, respectively.  In particular, Section \ref{sec:revopt} discusses several prominent deep learning architectures that focus on revenue maximization or cost minimization: RochetNet, RegretNet, MysersonNet,  MenuNet,  RegretFormer,  Budgeted RegretNet, and Stage-IC and Dynamic-IC. Sections \ref{sec:swmax}, \ref{sec:fair}, and \ref{sec:BudgetBalanced} then address the DL-based approach for welfare maximization, fairness of allocation, and budget balance, respectively.  Section \ref{sec:CaseStudies} demonstrates the efficacy of the DL-based approach through three illustrative case studies: (a) efficient energy management in a vehicular network; (b) resource allocation in a mobile network; and  (c) designing a volume discount procurement auction for agricultural inputs. Section \ref{sec:Conclusion} provides a summary and concludes the paper.

The paper by Zhang \cite{ZHANG21}
%, "A survey of online auction mechanism design using deep learning approaches," (arXiv:2110.06880, 2021) desired objectives 
is the lone existing survey paper in this area. The paper by Zhang only provides an overview of how deep learning techniques are being applied to the field of online auction mechanism design. The survey explores only a few deep learning methods for designing auctions  and serves only a limited purpose. Covers only papers until 2020. Our survey is comprehensive: (a) provides an overview of mechanism design with a detailed illustrative example; (b) covers four broad categories of these mechanisms: revenue maximizing (cost minimizing) mechanisms; welfare maximizing mechanisms; mechanisms with fairness in allocation; and mechanisms with budget balance; and (c) discusses three illustrative applications. 

%To enhance the readability of this paper, we have included Table 1 that describes the main symbols used in this paper and Table 2 which lists all the acronyms. 
% The symbols table is shifted to Section 3
% THe acronyms table has been dropped. It now appears after \end{document}

In our survey paper, we have attempted to explain the key concepts and details, drawing upon a number of papers in the current literature, which we have cited extensively in this paper. We have reproduced several illustrative pictures from the literature with an explicit citation to the source of the pictures; in all such cases, we have also obtained email permissions from the authors of these papers.

%%%%%%%%%%%%%%%%%%%%%%%%%%%%%%%%%%%%%%%%%
\section{Mechanism Design: An Overview}\label{sec:ClassicalResults}

%\subsection{Mechanism Design}
Mechanism design can be described as the art of designing the rules of a game to achieve a specific desired outcome \cite{NARAHARI14, MASCOLELL95}. The main focus of mechanism design is to create institutions or protocols that satisfy certain desired objectives, assuming that individual agents are rational and intelligent and will act strategically and may hold private information that is relevant to the decision at hand.
%MRB: Should we provide reference for explanation of rational, intelligent, etc.?
In a mechanism design setting, a policy maker (also called social planner or mechanism designer) faces the problem of aggregating the announced preferences of multiple agents into a collective (or social), system-wide decision when the actual preferences of the agents are not publicly known.  The essential technique that mechanism design uses to accomplish this is to
induce a game among the strategic agents in such a way that in an equilibrium of the induced game, the desired system-wide solution is implemented. Informally, a mechanism induces strategic players to do what the social planner would like them to do. Mechanism design has widespread and significant applications in modern web-based and AI systems \cite{BORGERS15, NISAN07, NARAHARI14, NARAHARI09}.

%In mechanism design, there exists a social planner (or policy maker or mechanism designer) and a set of players who are assumed to be rational and intelligent. The planner designs the rules and protocols of the game 
%MRB:Can remove following phrase for space
%and when the players play the game 
%according to the given rules and protocols, 
%a desired outcome may be achieved in an equilibrium of the game. In this setup, the preferences of the players are not known to the planner (i.e., player preferences are private information) and the planner has to aggregate this information to arrive at a collective decision.   
\subsection{Mechanisms}
The notion of mechanisms was first introduced by Leonid Hurwicz (Nobel Laureate in Economic Sciences in 2007) in 1960 \cite{NARAHARI14}. William Vickrey (Nobel Laureate in Economic Sciences in 1996) authored a seminal paper in 1961 \cite{VICKREY61} where he introduced the celebrated Vickrey auction (second price auction) and showed that the mechanism induces buyers in an auction to bid their true valuations without having to worry about what the other buyers bid. The Vickrey auction can be considered as a significant landmark in mechanism design theory. Edward Clarke \cite{CLARKE71} and Theodore Groves \cite{GROVES73} came up with generalizations of Vickrey mechanisms and paved the way for the famous VCG mechanisms (Vickrey-Clarke-Groves) that define a broad class of mechanisms satisfying certain very desirable properties. The discipline of mechanism design has made phenomenal advances in the decades following the 1960s and has found widespread applicability in a wide variety of applications and areas. The scope of mechanisms includes auctions and markets, trading institutions, regulation and auditing, social choice theory, resource allocation in multi-agent systems, electronic commerce, Internet advertisement, crowdsourcing, industrial procurement, spectrum allocation, etc. \cite{BORGERS15, NISAN07, NARAHARI14, NARAHARI09}. 

Mechanisms in which monetary transfers are allowed (for example, auctions) are known as {\em mechanisms with money\/} whereas mechanisms where there are no monetary transfers (such as voting mechanisms) are called {\em mechanisms without money\/} \cite{NARAHARI14}. In this paper, we are concerned with mechanisms with money. In particular, we focus on auctions, which are arguably the most popular example of mechanisms with money. Although our discussion is centered on auctions, all the results presented here can be generalized to all mechanisms with money.

\subsection{Auctions}
Auctions provide a natural example of mechanisms with money. An auction is a mechanism for assigning a set of goods to a set of agents based on bids announced by the agents. Auctions have been used for hundreds of years for trading of goods and services and their adoption in electronic commerce has risen exponentially in the past few decades. There is intense interest in the design of auctions due to their widespread applicability to traditional settings such as paintings and memorabilia as well as their popular adoption in more recent areas such as Internet advertising, keyword auctions on search engines, airport slot allocation, spectrum allocation, industrial procurement, and resource allocation in general.  A survey of auctions appears in several classic articles, including \cite{KLEMPERER99, MCAFEE87}. 

In an auction (and, in general, in mechanisms with money), the mechanism can be decomposed into an allocation rule and a payment rule. The allocation rule identifies the winners based on the bids of the agents while the payment rule computes the transfer of money among agents. In a forward (selling) auction where a seller is auctioning objects to potential buyers based on their bids, the allocation rule will determine which buyers win the objects and the payment rule determines how much they will pay to the seller.  In a reverse (procurement) auction, where a buyer is procuring objects from potential sellers based on their bids, the allocation rule will determine which sellers are selected to sell which objects and the payment rule determines how much money the winning sellers will receive.

\subsection{Desirable Properties of a Mechanism}\label{sec:DLMeetsAuctions}
We now provide a list of important properties that a mechanism is expected to satisfy. This list is taken from \cite{NARAHARI14}. As already stated, a mechanism induces a game among strategic agents in such a way that desirable properties are exhibited in an equilibrium of the game. The different equilibrium solutions that could be considered are dominant strategy equilibrium, Bayesian Nash equilibrium, and ex-post Nash equilibrium. For ease of understanding, we use the example of auctions while describing desirable properties.

\subsubsection*{Incentive Compatibility  (IC)} 
Incentive compatibility ensures truthful bidding by agents and is a fundamental requirement of any auction mechanism. The most powerful version of IC is dominant  strategy incentive compatibility (DSIC). DSIC ensures that, irrespective of the bids of the other consumers, it is in the best interest of each agent to bid their true value. On the other hand, Bayesian Nash incentive compatibility (BIC) only assures that bidding true valuation is a best response conditioned on all other agents bidding their true valuations. The VCG mechanisms remarkably satisfy DSIC.

\subsubsection*{Allocative Efficiency}
A mechanism is said to be allocatively efficient if the allocation rule maximizes the social utility (that is, the sum of utilities of all agents). This is also called social welfare maximization (SWM). The VCG mechanisms satisfy SWM as well.

\subsubsection*{Individual Rationality (IR)}
Individual rationality ensures that all the agents obtain non-negative utility by participating in the auction. The most powerful version of IR is ex-post IR, which implies that the utility to each participating player will be non-negative irrespective of the bids of the other players. Other forms of IR are interim IR and ex-ante IR \cite{NARAHARI14, MASCOLELL95}. 

\subsubsection*{Budget Balance}
An auction is said to be weakly budget balanced if, in all feasible outcomes, the payments by buyers will exceed or equal the receipts to the sellers. Strong budget balance is achieved if equality holds in the total payments and total receipts. Budget balance ensures that there is no need to infuse money into the system and the mechanism is sustainable.

\subsubsection*{Revenue Maximization}
%This means the expected total revenue generated for the farmers is maximized. FOPT implies a farmer-friendly auction.
It is desirable, in a forward auction, that the expected total revenue generated for the seller(s) is maximized. Often, rather than revenue maximization, the goal of the seller(s) could be profit maximization, which is revenue minus production cost.

\subsubsection*{Cost Minimization}
In a reverse auction or procurement auction, the buyer(s) would be interested in minimizing the expected total cost of procurement.
%This means the expected total cost to the consumers is minimized. COPT implies a consumer-friendly auction.
%For running multiple successful iterations of the mechanism, it is imperative that the expected total cost incurred by the consumers is minimized. COPT implies a consumer-friendly auction.

%\subsubsection*{Business Constraints (BUS)}
%It would help the FC run the auctions effectively over a longer time duration if certain business rules were implemented. 
%Business constraints refer to constraints such as having a minimum number of winning consumers (to avoid monopoly by a single or small number of consumers),  a maximum number of winning consumers (to minimize logistics costs), a maximum fraction of business to be awarded to any consumer, etc. 

\subsubsection*{Fairness of Allocation}
Fairness implies that the winning agents are chosen in a fair way. An index of fairness is envy-freeness -- no agent can increase their utility by adopting another agent's outcome. If envy-freeness is not achievable, the next best option is envy minimization (which is what we pursue in this paper).
%It should be noted that Nash Social Welfare and Business Constraints also, in their own respective ways, promote fairness to a certain extent. However, envy minimization is a popular, exclusive  measure of fairness.

%VR: Below para is almost completely lifted from previous paper.
%%%% YN - I have edited this
%%%%%%% Content existing here is now shifted after \end{document}
\subsection{Possibilities and Impossibilities in Mechanism Design}
The trade-offs involved in designing mechanisms with desirable properties such as incentive compatibility, individual rationality, social welfare maximization, and budget balance are quite complex and intricate \cite{NARAHARI14, MASCOLELL95}. There is vast literature in mechanism design theory that comprises possibility results and impossibility theorems.
As an example, VCG mechanisms satisfy dominant strategy incentive compatability, social welfare maximization, weak budget balance, and individual rationality. They do not satisfy strict budget balance. In fact, it is impossible for any mechanism to satisfy   dominant strategy incentive compatability, social welfare maximization, and strong budget balance.  Similarly, no mechanism can satisfy incentive compatability, social welfare maximization, strong budget balance, and individual rationality. When we include fairness of allocation and revenue maximization (or cost minimization), the situation gets worse and most subsets of these properties become infeasible.

In different practical applications, different subsets of these properties become desirable. For example, in a procurement auction, the buyer would like to minimize the total cost of procurement and would like the mechanism to be incentive compatible and individually rational, in addition to assuring a fair allocation as well as satisfying some business constraints. This is clearly in the realm of impossibility. Such an example in the context of agriculture inputs procurement is discussed  in Section \ref{subsec:AgriCaseStudy}. As a second example, a market mechanism for matching buyers with sellers should satisfy incentive compatibility, individual rationality, budget balance, and fairness of allocation, which again is in the realm of impossibility. Mechanism design researchers have investigated numerous such impossible situations and adopted a wide variety of approaches to circumvent these impossibilities. In this paper, we survey the various  deep learning based methods that have been developed to design a mechanism that satisfies the required subset of properties in a maximal or approximate way, by minimizing an appropriately crafted loss function. Numerous practical applications will benefit from this approach as shown in Section \ref{sec:CaseStudies}.

%%%%%%%%%%%%%%%%%%%%%%%%%%%%%%%%%%%%%%%%%%%
%%%%%%%%%%%New Section 
%%%%%%%%%%%%%%%%%%%%%%%%%%%%%%%%%%%%%%%%%%%%
\section{An Illustrative Example: Learning a Revenue Maximizing Auction}\label{sec:DLMDExample}
In this  section, we provide the example of an auction mechanism which is a generalization of the famous Myerson auction \cite{MYERSON81} for selling a single indivisible item. Such auctions are relevant in a wide variety of resource allocation settings. While the Myserson auction maximizes the expected revenue subject to Bayesian incentive compatibility and interim individual rationality, the extension we discuss does not satisfy these properties simultaneously and we show how the deep learning based approach is used to tackle this problem.

\subsection{Initial Setup and Notation}
Consider a set of buyers $N=\{1,2,...,n\}$  and a single seller with a set of items to sell, $M=\{1,2,..,m\}$. We use $i,j$ for buyers and $k,l$ for items. Table \ref{tab:notations} summarizes the notation used in the rest of the paper. Let $v_{i,l}$ be the valuation of the $i^{th}$ buyer for $l^{th}$ item for all $i \in N$ and $l \in M$. Also, let the valuation vector of buyer $i$ be, $\textbf{v}_i:2^M \rightarrow R_{\ge 0}$ such that 
 \begin{itemize}
     \item $x_i \le \textbf{v}_i(S)=\sum_{l \in S} v_{i,l} \le y_i$ for $i^{th}$ buyer having additive valuation
     \item $x_i \le \textbf{v}_i(S)=\max_{l \in S} v_{i,l} \le y_i$ for $i^{th}$ buyer having unit demand valuation
 \end{itemize}
where $\textbf{v}_i(S)$ is the valuation of the $i^{th}$ buyer for set $S \subseteq M$. %$\forall S$. 

\begin{table}[htb]
    \centering
    \begin{tabular}{|c|c|}
        \hline
        Notation & Description \\
        \hline
        $N$ & set of $n$ agents \\%buyers/sellers
        $M$ & set of $m$ items \\
        $i,j$ & index of agent\\%buyer/seller
        $k,l$ & index of item \\
        $v_{i,l}$ & valuation of $i^{th}$ agent for $l^{th}$ item \\
        $\textbf{v}_i$ & valuation vector of agent $i$ \\
        $\textbf{b}_i$ & bid vector of agent $i$ \\
        $x_i,y_i$ &  lower and upper bounds for $\textbf{v}_i$ \\
        $f_i$ & probability density function:  $[x_i,y_i] \rightarrow R_{+}$ \\
        $F_i$ & cumulative  distribution function: $[x_i,y_i] \rightarrow [0,1]$ \\
        $\textbf{V}_i$ & set of valuations with probability distribution $F_i$ \\
        $\textbf{v}$ & valuation profile $(\textbf{v}_1,...,\textbf{v}_n)$ \\
        $\textbf{V}$ & set of valuation profiles $\prod_{i=1}^n \textbf{V}_i$ \\
        $\textbf{P}$ & allocation rule defined as $\textbf{V} \rightarrow [0,1]^{m\text{x}n}$ \\
        $\textbf{t}$ & payment rule defined as $\textbf{V} \rightarrow R^{n}_{\ge 0}$ \\
        $u_i$ & utility of agent $i$ \\
        $U_i$ & expected utility of agent $i$\\
        $\textbf{w},\textbf{w}_0$ & parameters of thee neural network \\
        $B_i$ & payment budget for agent $i$ \\
        $R_i$ & regret budget for agent $i$ \\
        $T_i$ & type space for agent $i$ \\
        $W$ & public value summarization function \\
        \hline
    \end{tabular}
    \caption{Notation and symbols}
    \label{tab:notations}
\end{table}

Generally, a buyer can give their preference of items by bidding $\textbf{b}_i \le \textbf{v}_i$ along with a unit vector that specifies the required items. We assume that every buyer $i$ selects a valuation $\textbf{v}_i$ randomly from a set of valuations $\textbf{V}_i$ that has probability distribution $F_i:[x_i,y_i]\rightarrow[0,1]$ and Probability density function $f_i:[x_i,y_i]\rightarrow R_+$. We denote the valuation profile as $\textbf{v} = (\textbf{v}_1,\dots ,\textbf{v}_n)$ and the set of valuation profiles as $\textbf{V} = \Pi_{i=1}^n \textbf{V}_i$ where $\textbf{v} \in \textbf{V}$. We further denote $\textbf{v}_{-i}$ as the valuation profile $\textbf{v} = (\textbf{v}_1,\dots,\textbf{v}_n)$ without valuation $\textbf{v}_i$, (similarly for $\textbf{b}_{-i}$) and $\textbf{V}_{-i} = \Pi_{j \ne i} \textbf{V}_j$ as the set of possible valuation profiles. The seller knows the probability distribution $F = (F_1,\dots,F_n)$, but does not know the true valuations of the buyers. %Similarly, each buyer $i$ does not know valuations of other buyers but knows only probability distributions $(F_j)_{ \forall j \ne i}$ and knows his true valuation. 

Similarly, each buyer $i$ would know, in addition to their own true valuation, only the probability distributions $(F_j)_{ \forall j \ne i}$ and not the true valuations of the other buyers.

We also assume that the valuation of all buyers are statistically independent random variables. %and hence the density function is $f(\textbf{v}) = \Pi_{i \in N} f_i(\textbf{v}_i)$. 

With this given setup, a seller has to design an auction $(\textbf{P},\textbf{t})$, where $\textbf{P}$ denotes the allocation rule and $\textbf{t}$ denotes the payment rule. The allocation rule (also known as winner determination rule) is defined as $\textbf{P}:\textbf{V}\rightarrow [0,1]^{m\times n}$ with the probability of allocation matrix $\textbf{P}(\textbf{v}) = [p_{i,l}(\textbf{v})]_{i \in N,l\in M}$. Here $p_{i,l}(\textbf{v})$ is the probability of allocating $l^{th}$ item to the $i^{th}$ buyer for the given valuation profile $\textbf{v} \in \textbf{V}$ that satisfy the following conditions.
 \begin{equation} \label{eqn:prb1}
     0 \le p_{i,l}(\textbf{v}) \le 1 \quad \forall i\in N,l\in M,
\end{equation}
\begin{equation}\label{eqn:prb2}
     \sum_{i \in N} p_{i,l}(\textbf{v}) \le 1 \quad \forall l \in M.
 \end{equation}
The payment rule is defined as a function $\textbf{t}:\textbf{V} \rightarrow R_{\ge 0}^n$, with $\textbf{t}(\textbf{b}) = [t_i(\textbf{b})]_{i \in N}$, where the seller charges $t_i(\textbf{b})$ to buyer $i$ for given bid profile $\textbf{b} \in \textbf{V}$. The utility function of buyer $i$ is given by

\begin{equation}\label{eqn:uti_bidder}
    u_i(\textbf{P}(\textbf{b}),\textbf{t}(\textbf{b});\textbf{v}_i) = \sum_{l \in M}p_{i,l}(\textbf{b}) v_{i,l} - t_i(\textbf{b}) \quad \forall i \in N 
\end{equation}

and the expected utility is 

\begin{equation}\label{eqn:expected_uti_bidder}
U_i(\textbf{P}(\textbf{b}),\textbf{t}(\textbf{b});\textbf{v}_i) = E_{\textbf{V}_{-i} \sim F_{-i}} [u_i(\textbf{P}(\textbf{b}),\textbf{t}(\textbf{b});\textbf{v}_i)]
\end{equation}
The revenue of the seller is given by

\begin{equation}\label{eqn:uti_seller}
    u_0(\textbf{b}) = \sum_{i \in N} t_i(\textbf{b})
\end{equation}
and the seller's expected revenue is 

\begin{equation}\label{eqn:expected_rev_seller}
U_0(\textbf{v}) = E_{\textbf{b} \sim F} [u_0(\textbf{b})]
\end{equation}

For a feasible auction mechanism, the following conditions have to be satisfied in addition to Eqs. \ref{eqn:prb1} and \ref{eqn:prb2}. \newline

\begin{itemize}
    \item \textit{Individual Rationality}:\\ $U_i(\textbf{P},\textbf{t};\textbf{v}_i) \ge 0 \quad \forall i \in N \;  \forall \textbf{v}_i \in \textbf{V}_i$. %MRB:\\ A stronger version would be ex-post IR: u_i() \geq 0...
    \item \textit{Incentive Compatibility}: The buyer has to report truthfully. i.e., if the buyer bids $\textbf{v}'_i$ when the buyer's true valuation is $\textbf{v}_i$, then \\
    $U_i(\textbf{P}(\textbf{v}_i,\textbf{b}_{-i}),\textbf{t}(\textbf{v}_i,\textbf{b}_{-i});\textbf{v}_i) \ge \\     
    U_i(\textbf{P}(\textbf{v}'_i,\textbf{b}_{-i}),\textbf{t}(\textbf{v}'_i,\textbf{b}_{-i});\textbf{v}'_i) \\ \forall i \in N$ and  $\textbf{v}_i,\textbf{b}_i \in \textbf{V}_i$ and $\textbf{b}_{-i} \in \textbf{V}_{-i}$. \newline
\end{itemize}

The main goal of a feasible auction design problem is to find the probability of allocation matrix $\textbf{P}$ and payment rule $\textbf{t}$ to maximize the expected utility of the seller $U_0(\textbf{v})$ (Eq. \ref{eqn:expected_rev_seller}).

\subsection{Ad-Auction: Problem Formulation}
As an instance of the above auction, let us consider an online ad-auction where a website would like to sell time slots for advertising products to two firms ($n=2$ and $m=2$). %$n=2$ items (time slots) and $m=2$ buyers. 
Let the valuation of buyer $i$ for item $k$ be $v_{i,k}$ for $i=1,2$ and $k=1,2$ derived from Uniform Distribution $U[0,1]$. Also, assume buyers are truthful and have additive valuation. That is $\textbf{v}_i = v_{i,1} + v_{i,2}$ and has triangular probability density function $f_i(\textbf{v}_i) = \Lambda[0,2]$ for $i=1,2$. %MRB:We don't have an appendix(\textcolor{black}{refer appendix}). 
Assume that the seller will compute the allocation probability for an item $k$ to buyer $i$ as $p_{i,k}$ and the total amount payable by buyer $i$ as $t_i$ for a given valuation vector $\textbf{v} = [\textbf{v}_1,\textbf{v}_2]$. %where $\textbf{v}_1 \in \textbf{V}_1, \textbf{v}_2 \in \textbf{V}_2$ and $\textbf{v} \in \textbf{V} = \textbf{V}_1 \times \textbf{V}_2$. 
The buyer $2$ does not know $\textbf{v}_1$ but knows only the probability density function $f_1(\textbf{v}_1)$ and vice versa. Then, utility of buyers 1 and 2 are given as

\begin{equation}\label{eq:utiuser1}
    u_1(\textbf{P}(\textbf{v}),\textbf{t}(\textbf{v});\textbf{v}_1) = p_{1,1}(\textbf{v})v_{1,1} + p_{1,2}(\textbf{v})v_{1,2} - t_1(\textbf{v})
\end{equation}
\begin{equation}\label{eq:utiuser2}
    u_2(\textbf{P}(\textbf{v}),\textbf{t}(\textbf{v});\textbf{v}_2) = p_{2,1}(\textbf{v})v_{2,1} + p_{2,2}(\textbf{v})v_{\textbf{v}2,2} - t_2(\textbf{v})
\end{equation}
The expected utility of buyer $1$ is 
\begin{equation}\label{eq:Exputiuser1}
    U_1(\textbf{P}(\textbf{v}),\textbf{t}(\textbf{v});\textbf{v}_1) = \int_{R} u_1(\textbf{P}(\textbf{v}_1,\textbf{v}_2),\textbf{t}(\textbf{v}_1,\textbf{v}_2);\textbf{v}_1) f_2(\textbf{v}_2) d\textbf{v}_2
\end{equation}
Similarly, the expected utility of buyer $2$ is
\begin{equation}\label{eq:Exputiuser2}
    U_2(\textbf{P}(\textbf{v}),\textbf{t}(\textbf{v});\textbf{v}_2) = \int_{R} u_2(\textbf{P}(\textbf{v}_1,\textbf{v}_2),\textbf{t}(\textbf{v}_1,\textbf{v}_2);\textbf{v}_2) f_1(\textbf{v}_1) d\textbf{v}_1
\end{equation}
The utility/revenue of the seller is given by
\begin{equation}\label{eq:utiseller}
    u_0(\textbf{v}) = t_1(\textbf{v}) + t_2(\textbf{v})
\end{equation}
and an expected utility of the seller is
\begin{equation}\label{eq:Exputiseller}
    U_0(\textbf{v}) = \int_{R}\int_{R}  u_0(\textbf{v}_1,\textbf{v}_2)f_1(\textbf{v}_1)f_2(\textbf{v}_2)d\textbf{v}_1d\textbf{v}_2
\end{equation}

The seller has to solve the following optimization problem to compute the allocation probability and payments with IC assumption (known as revenue optimal auction with DSIC)
\begin{equation} \label{eq:optobj}
    \max_{p_{i,k},t_i, i=1,2 \& k=1,2} U_0(\textbf{v})
\end{equation}
such that
\begin{equation} \label{eq:optcon1}
    U_1(\textbf{P}(\textbf{v}),\textbf{t}(\textbf{v});\textbf{v}_1) \ge 0
\end{equation}
\begin{equation} \label{eq:optcon2}
    U_2(\textbf{P}(\textbf{v}),\textbf{t}(\textbf{v});\textbf{v}_2) \ge 0
\end{equation}
Eqs. \ref{eq:optcon1},\ref{eq:optcon2} are for satisfying IR constraints. 
% The following challenges are involved in designing an auction using traditional methods. 

% \subsection{Challenges Involved}

The main objective of the seller is to aggregate the bids from various buyers to compute the allocation matrix of items and the payment to the seller by the different buyers. The allocation matrix is a search problem and its complexity increases with the increase in number of buyers and number of items to sell. As given in the illustrative example, the seller has to compute the allocation matrix and payment vector by solving the optimization problem given in Eqs. \ref{eq:optobj}-\ref{eq:optcon2}. 
We consider that buyers draw their valuation from independent and identical probability distributions. The following challenges are involved in solving the given optimization problem. \newline

%MRB:Consider rephrasing parts of this section
\subsubsection*{Computational Complexity}
Solving the given optimization problem (Eqs. \ref{eq:optobj}-\ref{eq:optcon2}) for a given simple set up with $2$ buyers drawing valuations from independent and identical distributions 
%Independent and Identical distributions (IIDs) 
and a seller with $2$ items with IC as a condition is an NP hard problem. It is because, when the buyer draws valuation of each item from independent uniform distributions, the distribution of the valuation vector for that buyer is a triangular distribution for the case of additive valuations and it is the square of uniform distribution for the case of unit demand valuations. It is observed that there are no closed form expressions to compute expected utilities of buyers given in Eqs. \ref{eq:Exputiuser1} and \ref{eq:Exputiuser2} and for the seller given in Eq. \ref{eq:Exputiseller} for this given simple setup. Even for the case of a single buyer and seller in a multiple item setting, we can approximate the valuation set probability distribution to multivariate Gaussian distribution as per the Central Limit Theorem as the number of items increase in the case of additive valuations. But, there is no closed form expression to compute the utility of the seller and the solution to obtain the allocation probability and payments is an NP hard problem that satisfies IC and IR constraints. If we consider the valuation of each item of the buyer to be drawn from a distribution other than the uniform distribution, the computation of the overall probability distribution of valuation is also complex for the case of additive valuations. This leads to solving the optimization problem given above, which is NP hard. This similar result can be observed if the number of buyers are increased and the number of items are also increased.

Further, designing an auction mechanism that satisfies all the desirable properties simultaneously is a challenging task and it fails in certain scenarios that are described by impossibility results \cite{GIBBARD73},\cite{ARROW12}. \newline

\subsubsection*{Communication Complexity}
The revelation principle \cite{MCAFEE87} considers one shot revelation of true valuations of buyers to the seller by designing an IC mechanism. But it is not the case when players are selfish and may not be truthful for their personal benefits. In this scenario, a seller may be able to extract truthful valuations via multiple communications with the buyers. As opposed to the design of optimal auctions with unlimited communication \cite{MYERSON81}, the authors of \cite{MOOKHERJEE14} propose a feasible auction design that maximizes the utility subject to communication feasibility constraints. Here communication feasibility specifies either delay or cost of sending the bid in terms of number of bits. As the number of items $m$ and the number of buyers $n$ increases, the cost of communicating bids or delay in communicating bids also increases. It may also take more iterative communication cycles to extract truthful information about the valuations from the buyers. The authors obtained strong results for designing optimal decentralized decision making under the absence of trade-offs between incentives and information efficiency. In \cite{ZANDT07}, the authors study the relation between IC constraints and communication constraints. They also show that there exists a separation between checking IC constraints for Social Choice Function and the design of communication efficient protocol to compute the Social Choice Function for 
%MRB:Correct/explain phrase below
ex post Nash Implementation. \newline

\begin{figure*}[h]
    \centering
    \includegraphics[scale=0.6]{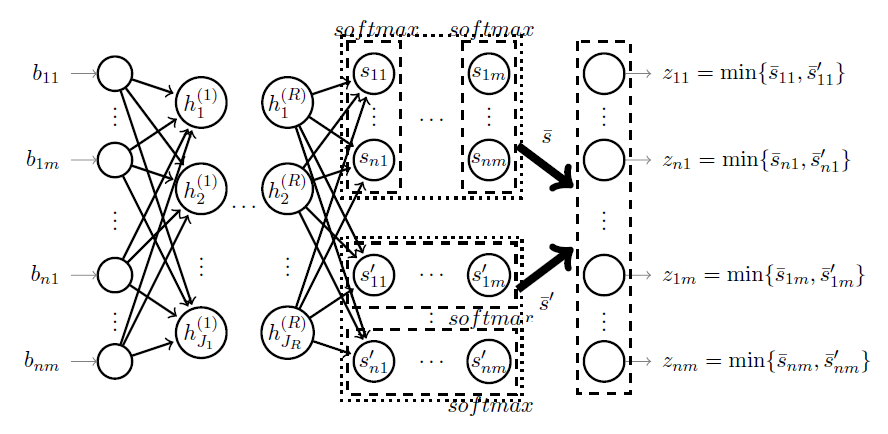}
    \caption{RegretNet: The allocation network for $n$ unit demand buyers with $m$ items (Figure reproduced from \cite{DUTTING23})}
    \label{fig:regretnetunitdemand}
\end{figure*}

%MRB:In following sentence, clarify which problem can be solved and how it can be solved. We could also rephrase to say that this is a solution to the problem mentioned in the sentence (designing an auction mechanism that satisfies all the desirable properties simultaneously is a challenging task and it fails in certain scenarios that are described by impossibility results) just before this section(Computational Complexity). Might want to relocate this paragraph.
\textcolor{black}{We can} use the RegretNet architecture \cite{DUTTING23} to solve the $m$ items and $n$ unit demand buyers problem as shown in Fig. \ref{fig:regretnetunitdemand}. This architecture aims to maximize the expected revenue to the seller under IC. 
%VR: Removing below text.
% As we consider this architecture for the case of $2$ buyers and a seller having $1$ item to sell, the architecture reduces to a simplified form by letting $n=2, m=1$. From this architecture, it is easy to compute the allocation rule and the payment rule without being concerned about the probability distributions of the data. Moreover, the complexity is significantly reduced. 
%MRB:Might want to explain how complexity is reduced
Further details about the architecture are provided in Section \ref{sec:revopt}.

%\section{A Summary of Approaches to Deep Learning Based Auction Design}\label{sec:SummaryOfDLApproaches}
%\section{A Summary of the Types of Auctions Designed using Deep Learning}\label{sec:SummaryOfDLApproaches}

%%%%%%%%%%%%%%%%%%%%%%%%%%%%%%%%%%%%
%%%%%%%%%%%%%%%  New section
%%%%%%%%%%%%%%%%%%%%%%%%%%%%%%%%%%%%
\section{Learning Revenue Optimal Mechanisms}\label{sec:revopt}

Theoretically, it has been proven that it is impossible to design revenue maximizing and incentive compatible auctions. In the following papers, the authors design neural networks that aim to maximize revenue while minimizing the violation of IC or satisfy IC while maximizing revenue.

In \cite{DUTTING23}, the authors propose an optimal auction design via RochetNet and RegretNet architectures that satisfy Dominant Strategy Incentive Compatibility (DSIC) and IR. The RochetNet architecture is based on constraining the search space so as to achieve DSIC and in this method the DSIC characterization is encoded within the Neural Network (NN) architecture. This architecture is suitable for a single buyer seeking to purchase multiple items from the auctioneer. In the case of the RegretNet architecture, approximate DSIC is achieved by differential approximation and Lagrangian augmentation. This method entails searching throughout a larger parameter space and is suitable for multi-buyer and multiple-item settings. The authors of \cite{FENG18} propose a DL based revenue optimal auction with private budget constraints of the buyers. This work is an extension of the RegretNet framework proposed in \cite{DUTTING23} that can handle both Bayesian IC and Conditional IC constraints. The authors in \cite{DENG20} introduce a new metric that quantifies Incentive Compatibility in auctions for both static and dynamic environments. This metric is data driven and is evaluated on the ad auction data. 

The authors in \cite{CURRY22} design revenue maximizing auctions while approximating incentive compatibility. They build on the RegretNet architecture by introducing a differentiable matching approach in the allocation segment. The RegretNet architecture cannot handle exactly-$k$ allocations, where each buyer is allocated exactly $k$ items in the auction. The primary difference between their architecture and the RegretNet architecture lies in the allocation network. They equate the allocation problem to finding a minimum cost bipartite matching between buyers and items, for a given cost matrix. The output of the allocation network is taken as this cost matrix, the matching problem for which is solved using the Sinkhorn algorithm which takes into consideration marginal constraints such as how many items a buyer must receive. The output of this algorithm provides the probability of allocation of each item to each buyer.

\cite{RAHME22} designs a neural network architecture known as EquivarianceNet, which aims to maximize revenue while limiting DSIC violations. It overcomes three limitations of the RegretNet architecture. The first is that RegretNet cannot produce symmetric mechanisms when the optimal solution is known to be symmetric. %Equivariance overcomes this by always producing symmetric auctions as the output. 
The second limitation of RegretNet is that it requires a large number of samples to converge. EquivarianceNet makes use of domain knowledge and limits the sample space so that it can converge with fewer samples. The third limitation of RegretNet is that it cannot perform inference when the number of buyers and items is different to what it was trained on. This is overcome through the use of exchangeable matrix layers as introduced in \cite{HARTFORD18}. The limitation in the EquivarianceNet architecture is that it assumes all buyers and items are permutation equivariant.

%In \cite{DUAN24} the authors design 
\cite{DUAN24} designs a revenue maximizing auction using neural networks that are inherently Dominant Strategy Incentive Compatible (DSIC) and Individually Rational (IR). Architectures like RegretNet minimize the violation of DSIC but cannot guarantee it. The authors make use of Affine Maximizer Auctions (AMAs) which are DSIC and IR. The parameters of an AMA consist of weights for each buyer and boosts associated with each allocation. The neural network architecture designed consists of an encode layer which encodes the bids into a joint representation, a menu layer to construct the allocation menu, a weight layer which learns the weights of the AMA, and a boost layer which learns the boost values of the AMA. As the mechanism is already DSIC, the only objective is to maximize the revenue.

\cite{LUONG18} proposes an optimal auction mechanism using a DL architecture for allocating edge computing resources to mobile users (i.e., miners) in a mobile blockchain environment. \cite{QIAN19} considers Mobile Network Virtual Operator (MNVO) that provides services via an auction based NN structure to allocate resources (subchannel and power) to its users. In both these applications, the authors portray the resource allocation problem as belonging to the single item multiple buyer setting and obtain optimal auctions in the sense that they maximize the revenue of the resource provider under DSIC and IR constraints. For this, bids from users are transformed via monotone transformation functions \cite{DUTTING23} and the second price auction with zero reserve price is used on the transformed bids to compute the allocation probability and payment rule.

The authors in \cite{ravindranath2024deepreinforcementlearningsequential} propose a revenue-optimal sequential combinatorial auction mechanism using a reinforcement learning framework. The authors introduce a fitted policy iteration approach to learn sequential auctions and implement a policy improvement step through the use of RochetNet or MenuNet. In \cite{liu2025interpretableautomatedmechanismdesign}, the authors propose an automated interpretable mechanism design using large language models and its application to both social welfare maximization  \& revenue maximization problems.

%\section{A Detailed Survey of the Types of Auctions Designed using Deep Learning}\label{sec:AnalyticalStudy}

%\subsection{\textbf{Revenue Maximizing Auctions}}
\subsection{RochetNet}

\begin{figure}[h]
    \centering
    \includegraphics[scale=0.6]{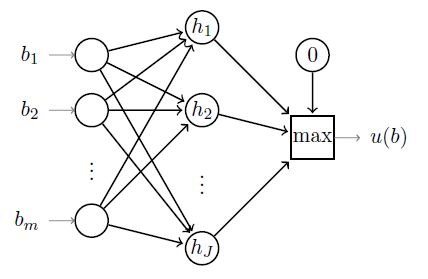}
    \caption{RochetNet: NN implementation of Eq. \ref{eq:rochet} (Figure reproduced from \cite{DUTTING23})} 
    \label{fig:rochet}
\end{figure}

\begin{figure}[h]
    \centering
    \includegraphics[scale=0.6]{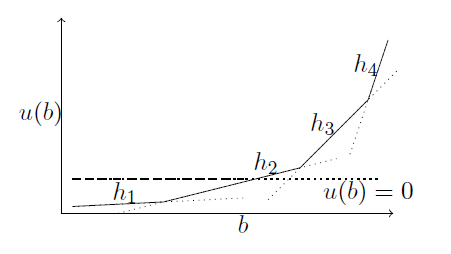}
    \caption{Example of utilization function generated from RochetNet (Figure reproduced from \cite{DUTTING23})}
    \label{fig:utirochet}
\end{figure}

An optimal auction design that maximizes the expected revenue to the seller under IC is proposed in \cite{DUTTING23}. \textcolor{black}{This approach} is a characterization based approach where IC constraints are encoded within the Neural Network (NN) architecture. The authors call such an NN a \textbf{\textit{RochetNet}}. This setup is for a single buyer with additive valuations and a seller with multiple items to sell. Let $\textbf{u}^{\textbf{w},\textbf{w}_0}(\textbf{v})$ be the seller's utility function, defined as

\begin{equation} \label{eq:rochet}
    \textbf{u}^{\textbf{w},\textbf{w}_0}(\textbf{v}) = \max\{\max_{k \in [J]} (\textbf{w}^T_k \textbf{v} + \textbf{w}_{0,k}), 0  \}
\end{equation}
where $(\textbf{w},\textbf{w}_0)$ are parameters and $\textbf{w}_k \in [0,1]^m, \textbf{w}_{0,k} \in R$. Then the utility defined in Eq. \ref{eq:rochet} is non-negative, monotonically non-decreasing, convex and $1-lipschitz$ w.r.t $l_1$-norm. 
The auction designed based on Eq. \ref{eq:rochet} is IC and IR. 
%MRB: Do we want to mention why it is IC and IR?
Each function $h_k(\textbf{v})= \textbf{w}^T_k \textbf{v} + \textbf{w}_{0,}$ is represented by a single neuron and a collection of such neurons will form one hidden layer in \textit{RochetNet} as shown in Fig. \ref{fig:rochet} and an example of a utilization function represented by it is shown in Fig. \ref{fig:utirochet}. 


The loss function considered for this NN optimization is the negated revenue of the seller with an approximate gradient given by

\begin{equation}\label{eq:rochetloss}
    \mathscr{L}(\textbf{w},\textbf{w}_0) = - E_{\textbf{V} \sim F } \{\sum_{k \in [J]} \textbf{w}_{0,k} \Tilde{\nabla}_k(\textbf{v})\}
\end{equation}

where,

\begin{gather*}
    \Tilde{\nabla}_k(\textbf{v}) = \text{Softmax}_k \{C.(\textbf{w}^T_1 \textbf{v} + \textbf{w}_{0,1}),\dots,C.(\textbf{w}^T_J \textbf{v} + \textbf{w}_{0,J})\},\\
    C>0
\end{gather*}

\subsection{RegretNet}

The next model in \cite{DUTTING23} uses differentiable approximation for IC constraints and the objective function is augmented with the term that measures how much IC constraints are violated. This approach is called a characterization free approach and is applicable to a mechanism where a seller with multiple items wishes to sell to multiple buyers. The authors define ex post regret for each buyer $i$ as:

\begin{equation} \label{eq:regret}
    rgt_i(\textbf{w}) = E_{\textbf{v} \sim \textbf{V}}\{\max_{\textbf{v}'_i \in \textbf{V}_i} u_i^{\textbf{w}((\textbf{v}'_i,\textbf{v}_{-i});\textbf{v}_i)} - u_i^{\textbf{w}((\textbf{v}_i,\textbf{v}_{-i});\textbf{v}_i)}\}
\end{equation}

Eq. \ref{eq:regret} specifies maximum increase in the utility of buyer $i$ with respect to all possible non-truthful bids. An auction mechanism is DSIC iff $rgt_i(\textbf{w}) = 0 \; \forall i \in N$ (except for zero event measures). The authors reformulated the problem (from Eq. \ref{eq:utiseller}) as

\begin{equation*}% \label{eq:utiregret}
    \min_{\textbf{w} \in R^d} E_{\textbf{v} \sim F}[-u_0(\textbf{v})]
\end{equation*}
such that
\begin{equation} \label{eq:consregret}
    rgt_i(\textbf{w}) = 0 \quad \forall i \in N
\end{equation}

\begin{figure*}[h]
    \centering
    \includegraphics[scale=0.5]{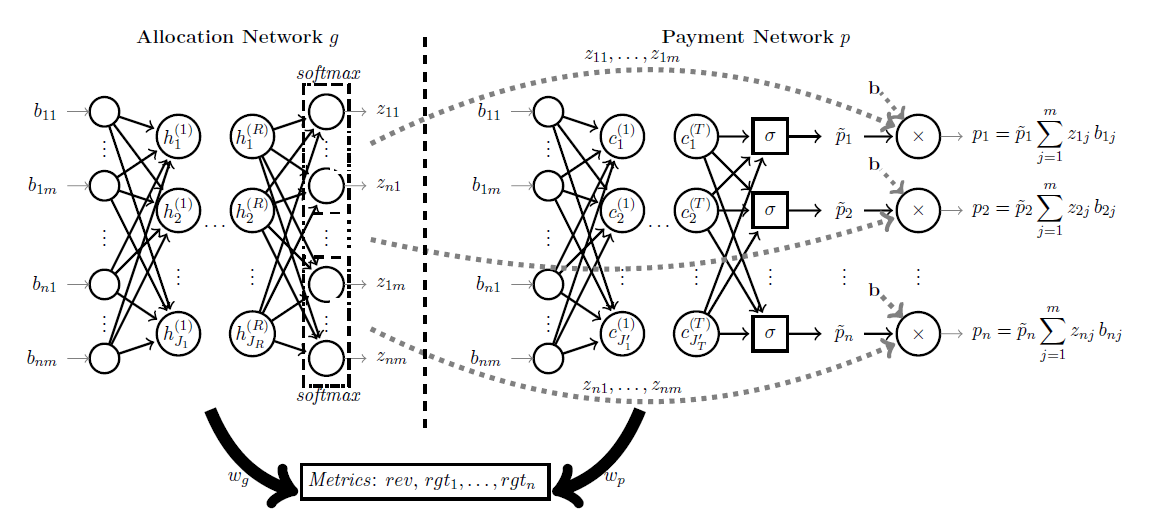}
    \caption{RegretNet: The allocation and payment networks for $n$ additive buyers with $m$ items (Figure reproduced from \cite{DUTTING23})}
    \label{fig:regretnet}
\end{figure*}

In addition to the above equations, the design has to satisfy IR constraints. From a given sample of a large number of evaluation profiles, an empirical ex post regret is computed for each buyer $i$ from Eq. \ref{eq:regret}. The problem is then to minimize the empirical loss (negated revenue) subject to empirical regret for all buyers %(from equations \ref{eq:utiregret}-\ref{eq:consregret}) 
(from equation \ref{eq:consregret}) 
and is implemented by using the \textbf{\textit{RegretNet}} approach. In this approach, an NN is trained using an augmented Lagrangian method to compute the allocation rule and payment rule separately. Correspondingly, the architecture consists of an allocation network and a payment network. The output of this trained network is used to compute regret and revenue of the auction mechanism. Fig. \ref{fig:regretnet} shows the RegretNet architecture that consists of the allocation network and the payment computation network for a seller having $m$ items to sell to $n$ additive buyers and Fig. \ref{fig:regretnetunitdemand} for the allocation network for $n$ unit demand buyers. 
%MRB: Consider removing the figure fig:regretnetunitdemand if pressed for space
The authors define the revenue and expected ex-post regrets as functions of the allocation and payment network parameters. For the case of unit demand buyers, the network considers random allocation rules such that its total allocation probability for each buyer is at most one and at the same time considers that no item is over allocated, i,e, probability of allocation of an item is at most one. 

\subsection{MyersonNet}

\cite{DUTTING23} also proposes an NN architecture for a seller with a single item (by letting $m=1$ in their proposed setting) and multiple buyers and extends the work to a general setting with multiple buyers and multiple items with combinatorial valuations. Below, we provide the NN architecture for a single item with multiple buyers. In this setup, the valuation function becomes $\textbf{v} = [v_1,\dots v_n]$ for $n$ buyers with $\textbf{b} \le \textbf{v}$.% Let $p_i$ be the probability of allocating an item to buyer $i$ where the seller gets $t_i$ as payment from the $i^{th}$ buyer. Then, the utility of the $i^{th}$ buyer for all $i=1,2,\dots n$ is
Then, the utility of the $i^{th}$ buyer is

\begin{equation}\label{eqn:utisingleitem}
    u_i(p_i(\textbf{v}),t_i(\textbf{v});v_i) = p_i(\textbf{v})v_i - t_i(\textbf{v})
\end{equation}
and the expected utility/revenue of the seller for a payment conditioned on allocation $p_i(\textbf{v}) t_i(\textbf{v})$ to buyer $i$ is

\begin{equation}
    u_0^i(\textbf{p},\textbf{t}) = E_{\textbf{v} \sim F} \{\sum_{i=1}^n p_i(\textbf{v}) t_i(\textbf{v})\}
\end{equation}
%MRB:Remove above equation if hard pressed for space

Consider a distribution function $F_i$ having probability density $f_i$, then the virtual valuation function is given by $\xi_i(v_i) = v_i - \frac{1-F(v_i)}{f(v_i)}$ and for monotonically non-decreasing $\xi_i$, the corresponding distribution $F_i$ (and density $f_i$) is regular. 
This implies that the virtual valuation, $\Bar{\Phi}_i = \xi_i$ for regular distributions $F_i \; \forall i=1,2,\dots,n$. %MRB:N or n? or just \in N?
If these virtual functions $\xi_i$ are monotonically increasing, then the optimal auction can be designed by first applying monotone transformations to bids and then feeding these virtual values to the second price auction with zero reserve price, leading to an output $(P^o,t^o)$ with allocation $P^o(\Bar{b})$ and payment $\Bar{\Phi}_i^{-1}(t^o_i(\Bar{b}))$. This auction mechanism is DSIC \cite{DUTTING23}, \cite{ROCHET87}. These virtual value functions $\Bar{\Phi}_i$ are modelled using two layer feed forward network with min and max over linear functions for $K$ groups of $J$ linear functions as shown below.

\begin{equation}\label{eq:mayermonotonefunctions}
   \Bar{\Phi}_i(b_i) = \min_{k\in [K]}\max_{k \in [J]} \{w^i_{k,j} b_i + \beta^i_{k,j}\}
\end{equation}

Each of the above linear functions are non-decreasing and hence $\Bar{\Phi}_i$ is also non-decreasing. We can also set $w^i_{k,j} = e^{\alpha^i_{k,j}}$ for a bounded range of parameters $\alpha^i_{i,j} \in [-B,B]$. With this representation, an inverse transform $\Bar{\Phi}^{-1}$ can be obtained from the forward transform parameters as given below.

\begin{equation*}
    \Bar{\Phi}^{-1}_i(y) = \max_{k\in [K]}\min_{k \in [J]} e^{-\alpha^i_{k,j}}\{y - \beta^i_{k,j}\}
\end{equation*}

\begin{figure}[h]
    \centering
    \includegraphics[scale=0.8]{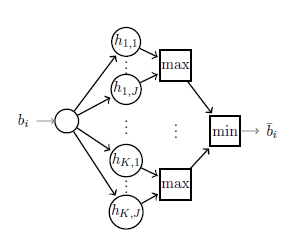}
    \caption{Monotone virtual value function implementation given in Eq. \ref{eq:mayermonotonefunctions} where $h_{k,j}(b_i) = e^{\alpha^i_{k,j}}b_i + \beta^i_{k,j}$ (Figure reproduced from \cite{DUTTING23})}
    \label{fig:monotonefunctions}
\end{figure}

\begin{figure}[h]
    \centering
    \includegraphics[scale=0.8]{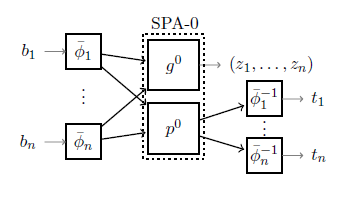}
    \caption{MyersonNet: monotone transformed bids are applied to Second Price auction network (Figure reproduced from \cite{DUTTING23})}
    \label{fig:myersonnet}
\end{figure}

An implementation of the monotone virtual value function given in Eq. \ref{eq:mayermonotonefunctions} using an NN is shown in Fig. \ref{fig:monotonefunctions} and application of monotone transformed bids to the Second Price Auction with zero reserve price (SPA-0) is shown in Fig. \ref{fig:myersonnet}.

\begin{figure*}[h]
    \centering
    \includegraphics[scale=0.8]{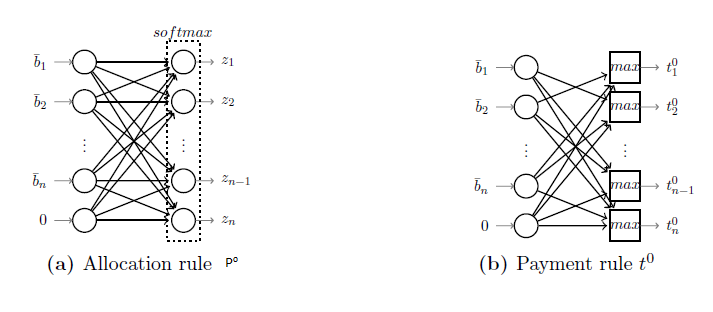}
    \caption{Second Price auction with zero reserve price network (Figure reproduced from \cite{DUTTING23})}
    \label{fig:spa0}
\end{figure*}

Fig. \ref{fig:spa0} models the second price auction with zero reserve price where the allocation rule is given in (a) and the payment rule is given in (b). It is observed that the allocation rule is approximated using a softmax activation function whereas payment rule is implemented using a maximum function as given below.

\begin{equation*}
    t^o_i(\Bar{b}) = \max\{\max_{j \ne i} \Bar{b}_j, 0\} \quad \forall i=1,2,\dots,n
\end{equation*}
%MRB:N? n?

%VR-done: Why does Figure 7 look blank in the middle?
%Ans: It is correct

In this work, the authors model IC and IR constraints as soft constraints by formulating the loss function as a linear combination of the revenue loss and the amount of violations of IC and IR constraints. The resulting auction mechanism may not be exactly IC. 

\subsection{MenuNet}

In \cite{SHEN18}, the authors propose an NN based optimal auction design by introducing another independent NN structure that captures the behaviour of buyers. This buyer network models the exact IC constraint. The authors propose an NN model for an auction setting where a seller with multiple heterogeneous ($m$ items) items would like to sell to a single buyer who may have an additive valuation function.

\begin{figure*}[h]
    \centering
    \includegraphics[scale=0.6]{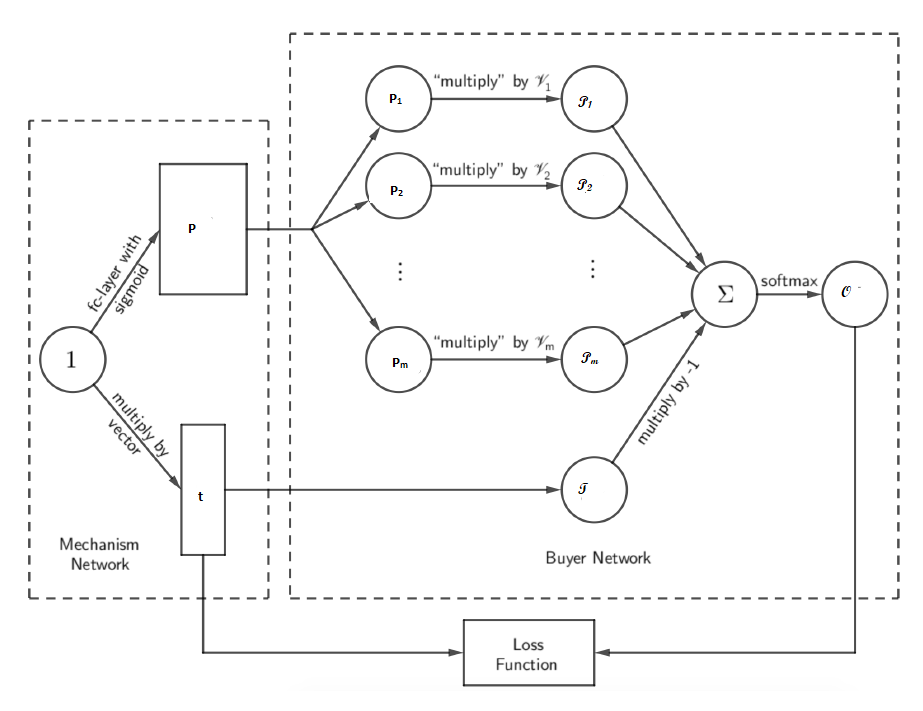}
    \caption{MENUNet Architecture (Figure reproduced from \cite{SHEN18})}
    \label{fig:menunet}
\end{figure*}

\begin{figure}[h]
    \centering
    \includegraphics[scale=0.4]{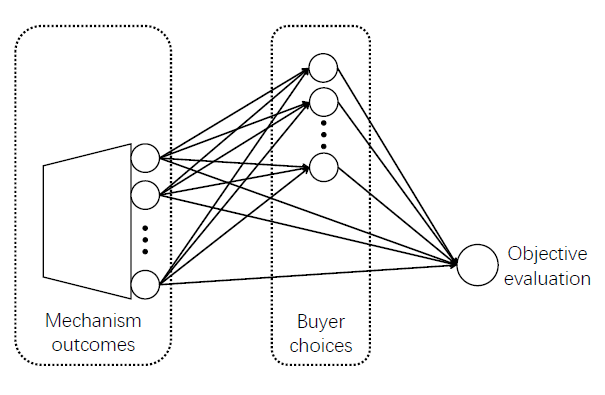}
    \caption{Naive Mechanism Architecture (Figure reproduced from \cite{SHEN18})}
    \label{fig:naivemechanism}
\end{figure}

\begin{figure}[h]
    \centering
    \includegraphics[scale=0.4]{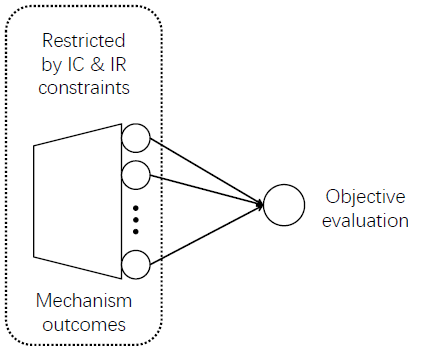}
    \caption{Direct Mechanism Architecture (Figure reproduced from \cite{SHEN18})}
    \label{fig:directmechanism}
\end{figure}

Let $\textbf{p}$ be the allocation vector and $\textbf{p}\in \mathscr{P} \subseteq [0,1]^m$ where $p_i$ is the probability of allocating an item $i$ to the buyer. Let $\textbf{v} = (v_1,\dots,v_m)$ be a value profile where $v_i \sim F_i$ is the value of an $i^{th}$ item with $0 \le v_i \le \hat{v}_i$ and let $V_i$ be the set of $d_i$ discrete values of $v_i$ in the interval $0 \le v_i \le \hat{v}_i$. Let us define $V = V_1 \times,\dots,\times V_m$, $\mathscr{F} = F_1 \times,\dots,\times F_m$ to be independent in each dimension and $t$ to be the payment. The proposed MenuNet architecture is shown in Fig. \ref{fig:menunet}. This architecture is based on the Naive mechanism (\cite{SHEN18}) which shows allocations along with corresponding items known as the menu and the buyer makes the most preferred choice. The high level structure of the Naive mechanism and the Direct mechanism is shown in Figs. \ref{fig:naivemechanism}, \ref{fig:directmechanism}. %MRB:Consider combining the 2 figures to save 1 line in caption
As shown in the figures, in the case of direct mechanisms, the objective is evaluated on the outcomes that are restricted by IC and IR constraints whereas in the case of Naive mechanisms, %the objective is computed from the buyer choices where its input is the mechanism's outcome. 
the mechanism's outcome serves as input to the buyer choices, which are used to compute the objective.

MenuNet consists of a mechanism network and a buyer network. The input of the mechanism network is a one dimensional constant $1$ and the outputs are an allocation matrix and a payment vector. The allocation matrix $P$ has rows of $m$ items and columns of $k$ menu items. Each column represents the allocation of all $m$ items. The payment vector $\textbf{t}$ represents the prices of $k$ menu items. In order to ensure IR of the buyer, the last column of $P$ and the last element of $\textbf{p}$ is set to $0$ which is a default choice for the buyer. A fully connected layer having a sigmoid function takes as input constant $1$ and forms each row $P_i$ of an allocation matrix, where $P_i$ represents allocation of $i^{th}$ item in all menu items. The payment vector $\textbf{t}$ is computed by multiplying the input constant by a scalar parameter. 

From Fig. \ref{fig:menunet}, it is observed that the tensor $\Psi_i$ = $d_1 \times \dots \times d_m $ is an $m$ dimensional tensor and its multiplication with $P_i$ forms the $m+1$ dimensional tensor $\mathscr{P}_i$ with size $d_1 \times \dots \times d_m  \times k$. It is easy to form the payment tensor $ \mathscr{T}$ with size $d_1 \times \dots \times d_m  \times k$. The utility is given by

 \begin{equation}
      \mathscr{O} =  \sum_{i=1}^m  \mathscr{P}_i -  \mathscr{T}
 \end{equation}
 
The loss function for a given strategy of the buyer $s(\textbf{v})$ for each value profile $\textbf{v}$ is given as

 \begin{equation}
      \mathscr{L} = -  \sum_{i=1}^m Prob[\textbf{v}] \textbf{t}^Ts(\textbf{v})
 \end{equation}
 
\subsection{RegretFormer}

\begin{figure*}[h]
    \centering
    \includegraphics[width=\textwidth,keepaspectratio]{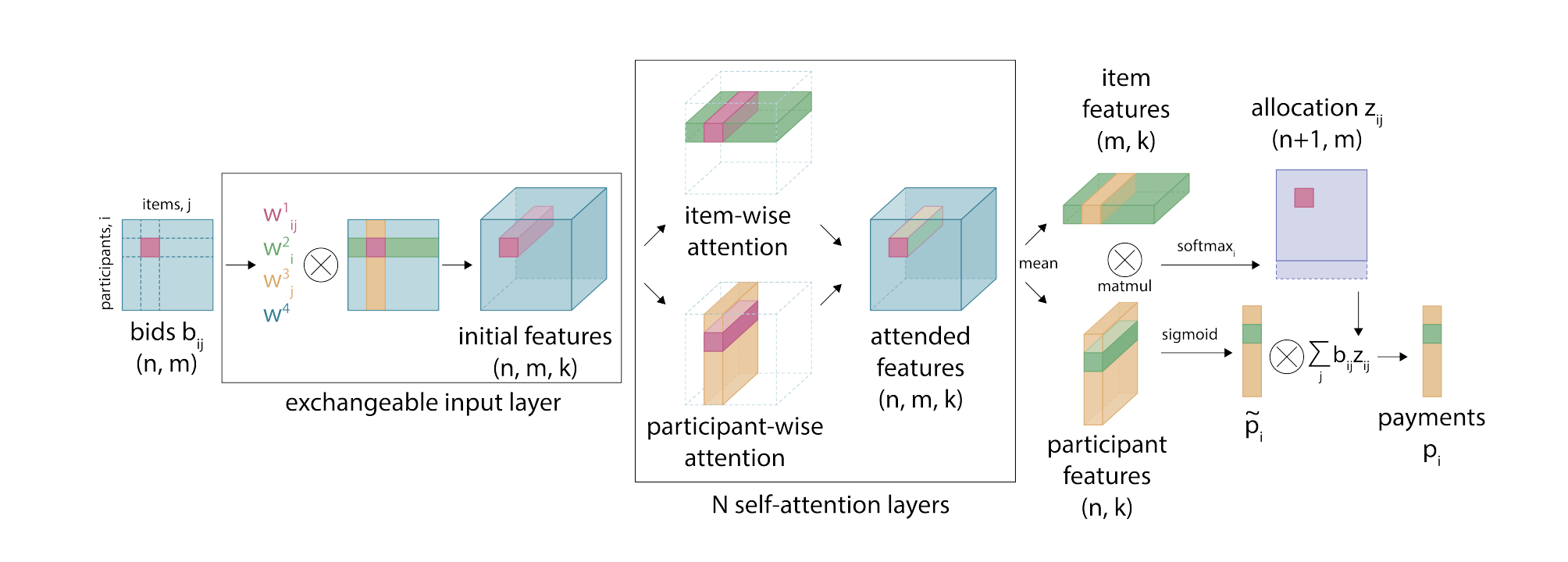}
    \caption{RegretFormer Architecture in (Figure reproduced from \cite{IVANOV22})}
    \label{fig:RegretFormer}
\end{figure*}

In \cite{IVANOV22}, the authors provide non-trivial extensions to the RegretNet architecture \cite{DUTTING23} in two ways. %MRB:Do we want to mention that the 2 ways are not mutually exclusive?
The first is, they propose a novel NN architecture based on attention layers which they call \textbf{\textit{RegretFormer}}, shown in Fig. \ref{fig:RegretFormer}. The RegretNet architecture has three main issues. It is not generalizeable to unseen number of items and buyers or to data with heterogeneous input sizes, it is sensitive to the order of items and buyers in the input bid matrix, and the fully connected layers alone do not produce the expressive power needed for auction design. RegretFormer tackles the latter 2 problems.%all these problems.

The second way they improve upon RegretNet is by attempting to maximize the revenue subject to a maximum regret cap. RegretNet aims to simultaneously maximize revenue and minimize regret by finding suitable values for the hyperparameters of the model. This technique imposes extremely precise hyperparameters for every experiment which have no predictability, i.e., selecting a set of values for the hyperparameters will not guarantee any outcome. This also means that whenever a new experiment must be conducted, for instance with a different number of items and buyers, hyperparameter tuning will need to be extensively performed to obtain accurate results. \cite{IVANOV22} maximizes the revenue given a maximum regret budget $R_{max}$ which is specified by the designer beforehand. This reformulates the problem in %Eqs. \ref{eq:utiregret},\ref{eq:consregret} as
Eq. \ref{eq:consregret} as

\begin{equation*}% \label{eq:utiregretformer}
    \min_{\textbf{w} \in R^d} E_{\textbf{v} \sim F}[-u_0(\textbf{v})]
\end{equation*}
such that

\begin{equation} \label{eq:consregretformer}
    E[rgt_i(\textbf{w})] \leq R_{max}
\end{equation}

%The dual of the Eqs. \ref{eq:utiregretformer},\ref{eq:consregretformer} 
The dual of Eq. \ref{eq:consregretformer} 
is minimized by introducing $\gamma$ as the Lagrangian multiplier, given as

\begin{equation} %\label{eq:utiregretformerdual}
    L_{outer}(\textbf{w}) = - \sum_{i \in N}P_i + \gamma \sum_{i \in N}\tilde{R_i}
\end{equation}

The dual gradient descent technique \cite{BOYD04} is applied by performing one gradient update for the parameters \textbf{w} and one gradient update for $\gamma$.
%The dual gradient descent technique \cite{BOYD04} is applied by performing a gradient update each for the parameters \textbf{w} and $\gamma$.

\subsection{Budgeted RegretNet}

There are scenarios where buyers have budget constraints in purchasing an item such as spectrum auctions, land auctions, etc. This constraint specifies that even though buyers know the exact value of the item, they are not able to pay that much money due to their financial constraints. This type of auction has the valuation of an item along with the budget constraint as private information of buyers. In \cite{FENG18}, the authors propose a DL based approach for maximizing revenue under the private budget constraints that satisfy both BIC and conditional IC constraints by extending the RegretNet framework \cite{DUTTING23} which they call Budgeted RegretNet. Let $T_i = (B_i, v_i) \sim F_i$ be the possible type space for buyer $i$, where $B_i$, $v_i$, are the payment budget and valuation for the $i^{th}$ buyer of a single item respectively (drawn from distribution $F_i$ and $\textbf{T} = (T_1,\dots, T_n) \sim F = \prod_{i=1}^nF_i$), and sold by a single seller where this information is private to buyer $i$. Every buyer bids with $t' \in T_i$ and the utility of each buyer $i$ given in equation \ref{eqn:utisingleitem}, changes to 

\begin{equation} \label{eqn:utiwithbc}
    u_i(T_i,t') = 
\begin{cases}
    p_i(t')v_i - t_i(t'),& \text{if } t_i(t') \le B_i\\
    -\infty,              & \text{otherwise}
\end{cases}
\end{equation} 

where $p_i(t')$ and $t_i(t')$ are the allocation and payment obtained from the NN.
We consider the budget constraint for buyer $i$, where  $i=1,\dots, n$ as 

\begin{equation} \label{eqn:bc}
    t_i(t') \le B_i \quad \forall t' \in \textbf{T}
\end{equation}

Eq. \ref{eqn:utiwithbc} is referred to as the utility of the buyer with budget constraints. If the budget constraint is violated, the buyer gets a higher penalty and the utility is considered as $-\infty$. 
%Similarly, a buyer may not report the actual budget and may report a budget less than that of the actual budget (i.e., $B'_i \le B_i$). 
%MRB:Confirm following para with Vishisht
A buyer may report a budget less than the actual budget (i.e., $B'_i \le B_i$). 
In this scenario, the corresponding DSIC can be termed as the Conditional DSIC (C-DSIC). For BIC, for given bid $t'$, we define an $i^{th}$ buyer interim allocation as $\mathscr{P}_i(t') = E_{T_{-i} \in F_{-i}}[p_i(t',T_{-i})]$ and interim payment as $\mathscr{T}_i(t')=E_{T_{-i} \in F_{-i}}[t_i(t',T_{-i})]$. The corresponding interim utility function is

\begin{equation} \label{eqn:interimutiwithbc}
    \mathscr{U}_i(T_i,t') = 
\begin{cases}
    \mathscr{P}_i(t')v_i - \mathscr{T}_i(t'),& \text{if } \mathscr{T}_i(t') \le B_i\\
    -\infty,              & \text{otherwise}
\end{cases}
\end{equation} 
and the interim budget constraint is

\begin{equation} \label{eqn:interimbc}
    \mathscr{T}_i(t') \le B_i \quad \forall t' \in \textbf{T}
\end{equation}
The corresponding Conditional BIC is 

\begin{equation} \label{conditionalBIC}
    \mathscr{U}_i(T_i,T_i) \ge \mathscr{U}_i(T_i,T'_i) \; \forall i=1,\dots, n \; \forall T_i \; T'_i \in \mathscr{T}_i
\end{equation}%MRB:N?

For a given auction $(\textbf{P},\textbf{t})$, the interim regret (maximum gain in interim utility by misreporting the buyer's choice) for the $i^{th}$ buyer is

%MRB:Check following equation. Also, check its overflow out of the margin
\begin{dmath}\label{eqn:interimregret}
    RGT_i(\textbf{P},\textbf{t}) = E_{T_i \sim F_i}[max_{T'_i \in \mathscr{T}_i} 1_{ \mathscr{T}_i(t') \le B_i} \mathscr{U}_i(T_i,T'_i) - \mathscr{U}_i(T_i,T_i)]
\end{dmath}
with indicator function $1_A$, where $A$ is either true or false. The IR penalty can be defined as
\begin{equation}\label{eqn:irpenalty}
    IRP_i(\textbf{P},\textbf{t}) = E_{T_i \sim F_i}[max(0, -\mathscr{U}_i(T_i,T_i))]
\end{equation}
and the budget constraint penalty can be defined as
\begin{equation}\label{eqn:bcp}
    BCP_i(\textbf{P},\textbf{t}) = E_{T_i \sim F_i}[max(0,  \mathscr{T}_i(T_i) - B_i)]
\end{equation}
The proposed NN architecture uses RegretNet, which maximizes the negative of expected revenue given by
\begin{equation}\label{eqn:rev}
    REV(\textbf{P},\textbf{t}) = - E_{t' \sim F}\{\sum_{i=1}^n  t_i(t')\}
\end{equation}
subject to Budget Constraints (BC), regret, and IR (given in Eqs. \ref{eqn:interimregret},\ref{eqn:irpenalty},\ref{eqn:bcp}) as
\begin{equation*}
    RGT_i(\textbf{P},\textbf{t}) = 0 \quad \forall i=1,\dots, n 
\end{equation*}
\begin{equation*}
    IRP_i(\textbf{P},\textbf{t}) = 0 \quad \forall i=1,\dots, n 
\end{equation*}
\begin{equation*}
    BCP_i(\textbf{P},\textbf{t}) = 0 \quad \forall i=1,\dots, n
\end{equation*}
It is also observed that the metrics given in Eqs. \ref{eqn:interimregret}, \ref{eqn:irpenalty}, \ref{eqn:bcp}, and \ref{eqn:rev} can be estimated from samples of type profiles $S = \{T^{(1)}, T^{(2)},\dots, T^{(L)}\}$. An unconstrained objective function can be formulated using an augmented Lagrangian approach and can be given as (without BC)
\begin{dmath} \label{eqn:legrangianwithoutbcp}
%\begin{align}
      \hat{\mathscr{L}}(\textbf{P},\textbf{t}) =  REV(\textbf{P},\textbf{t}) + \sum_{i=1}^n \lambda_{RGT,i} RGT_i(\textbf{P},\textbf{t})  \\ + \frac{\rho}{2} \sum_{i=1}^n RGT^2_i(\textbf{P},\textbf{t}) \\ +
                                                \sum_{i=1}^n \lambda_{IRP,i} IRP_i(\textbf{P},\textbf{t}) + \frac{\rho}{2} \sum_{i=1}^n IRP^2_i(\textbf{P},\textbf{t})
%\end{align}
\end{dmath}

The loss function with BC can be given as

\begin{dmath} \label{eqn:legrangianwithbcp}
    \mathscr{L}(\textbf{P},\textbf{t}) = \hat{\mathscr{L}}(\textbf{P},\textbf{t}) + \sum_{i=1}^n \lambda_{BCP,i} BCP_i(\textbf{P},\textbf{t}) \\ + \frac{\rho}{2} \sum_{i=1}^n BCP^2_i(\textbf{P},\textbf{t})
\end{dmath}

The proposed NN uses the loss function given in Eq. \ref{eqn:legrangianwithbcp} to train the model to compute the corresponding allocation and payment for each buyer. Here, $\{\lambda_{RGT,i}, \lambda_{IRP,i}, \\ \lambda_{BCP,i}\}$ are Lagrangian multipliers and fixed parameter $\rho >0 $ controls the weights on augmented Lagrangian.

Further, this work is extended to a multiple items auction setting. We observe that regret can be defined as a difference between utilities at best response and truthful bidding. Since regret has to be computed for every possible valuation in each step, its computational complexity is very high.

\subsection{Stage-IC and Dynamic-IC}

\cite{DENG20} proposes a new metric for quantifying IC, using a data driven computational approach. 
It is a measure of the marginal advantage of bidding a scaled version of the true value. 
This metric can be applied to both static and dynamic auctions. In dynamic auctions, agents' previous bids influence their future payments. The metric is described below.

Let us consider a dynamic auction (e.g., ad-auctions) with $n$ buyers and a single seller. It has a sequence of queries. For example in this type of ad-auctions, the winner (based on bid) will be allowed to display ads on the website. Let us assume that the $t^{th}$ bid arrives at time $t$ and let $v_t \in V = [0,1]$ be the valuation with respect to a single perspective buyer drawn independently from the distribution $F_t$ over V (where the corresponding probability density function is $f_t$). Also, assume that distributions at all stages are not necessarily identical. At each time step $t$, in response to the query, each buyer randomly chooses $v_t$ with distribution $F_t$ and bids $b_t \in V$ to the seller. The seller allocates the resource based on received bids from all buyers according to the allocation rule $p_t:V^t\rightarrow  [0,1]$ and payment rule $T_t:V^t\rightarrow R$, where $(p_t,T_t)$ is the dynamic auction outcome and $V^t$ is the first $t$ stage historical bids. Then, utility of a particular buyer at time $t$ is
\begin{equation*}
    U_t(b_{(1,t)};v_t) = v_t.p_t(b_{(1,t)}) - T_t(b_{(1,t)})
\end{equation*}
and with truthful bids for convergence at time $t$ (i.e., $b_t = v_t$), it is

\begin{equation*}
    \hat{U}_t(b_{(1,t)}) = b_t.p_t(b_{(1,t)}) - T_t(b_{(1,t)})
\end{equation*}
where $b_{(t^1,t^n)} = (b_{t^1},\dots,b_{t^n})$ represents the sequence of bids from time instants $t^1$ to $t^n$. In a dynamic auction, the objective of the buyer is to maximize their time discounted cumulative utility with respect to a discounting factor $\gamma \in [0,1]$. The optimal bidding strategy of the buyer at any time instant $t$ depends on future bidding strategies. 
Dynamic incentive compatibility (Dynamic-IC) implies that the buyer is assumed to bid truthfully at all subsequent time stages, and at all stages, Dynamic-IC holds. At any given time $t$, with bidding history $b_{(1,t-1)}$ and valuation $v_t$, for a dynamic auction $(p_t,T_t)$ to be Dynamic-IC we have

\begin{equation}\label{eqn:Dynamic-IC}
    v_t \in \argmaxA_{b_t} [ U_t(b_{(1,t-1)};v_t) +  \bar{U}_t(b_{(1,t-1)},b_t) ]
\end{equation}
where continuation utility $\bar{U}_t$ of a buyer is defined as the expected future utility under the assumption of truthful bidding and is given as

\begin{equation}\label{eqn:continuation_uti}
    \bar{U}_t(b_{(1,t)}) = \sum_{t'=t+1}^T \gamma^{t'-t} E_{v_{(t+1,t')}}\{\hat{U}_{t'}(b_{(1,t-1)},v_{(t+1,t')}) \}
\end{equation}

Stage IC can be achieved by substituting discounting factor $\gamma=0$ in Eq. \ref{eqn:continuation_uti}. The resulting equation obtained from Eq. \ref{eqn:Dynamic-IC} for any given time $t$, bidding history $b_{(1,t-1)}$ and valuation $v_t$ is

\begin{equation}
    v_t \in \argmaxA_{b_t} [ U_t(b_{(1,t-1)};v_t)]
\end{equation}

Below we provide definitions for Individual Stage-IC metric and Individual Dynamic-IC metric. For simplicity we remove dependency on time $t$ as it is for each time/stage and by definition it depends on previous bids. We also assume truthful bids by an individual at a given stage i.e., $b = v$

\begin{definition} (Individual Stage-IC metric)
    For a buyer with valuation distribution $F$ for a mechanism $(p,T)$ the Individual Stage-IC metric is given by
    
    \begin{equation*}
        I-SIC = \lim_{\alpha \rightarrow 0} \frac{E_{v\sim F}[\hat{U}((1+\alpha)v)]-E_{v\sim F}[\hat{U}((1-\alpha)v)]}{2\alpha E_{v\sim F}[vp(v)]}
    \end{equation*}
    where the numerator is the difference between expected utility when the buyer valuations are perturbed by $\alpha$ and $E_{v\sim F}[vp(v)]$ is the expected welfare of the buyer who bids truthfully. 
\end{definition}

\begin{definition}(Individual Dynamic-IC metric)
    For a given buyer valuation distribution  $F_{(1,T')}$ and discounting factor $\gamma$ for a dynamic mechanism $(p,T)$ the Individual Dynamic-IC metric for the first stage is given by
    
    \begin{equation*}
        I-DIC = \lim_{\alpha \rightarrow 0} \frac{\Delta s_1(\alpha) + \sum_{t'=2}^{T'} \gamma^{t'-1} \Delta d_{t'}(\alpha) }{2\alpha E_{v_1}[v_1p_1(v_1)]}
    \end{equation*}
    where $\Delta s_1(\alpha) = E_{v1}[\hat{U}_1((1+\alpha)v_1)-\hat{U}_1((1-\alpha)v_1)]$ and $\Delta d_{t'}(\alpha) = E_{v(1,t')}[\hat{U}_1((1+\alpha)v_1,v_{(2,t')})-\hat{U}_1((1-\alpha)v_1,v_{(2,t')})]$. These metrics (I-SIC and I-DIC) are computed using a data driven approach and are compared to these results' theoretical values. 
    %MRB:These results - which results? Need to explain clearly
\end{definition}

%%%%%%%%%%%%%%%%%%%%%%%%%%%%%%%%%%%%%%%%%%%%%%%%%%%%%%
%%%%%%%%%%%%%%%%%  New Section
%%%%%%%%%%%%%%%%%%%%%%%%%%%%%%%%%%%%%%%%%%%%%%%%%%%%%%
\section{Learning Welfare Maximizing Mechanisms}\label{sec:swmax}

%In \cite{TACCHETTI19} the authors focus on DL based approaches for auctions that are truthful, efficient, and minimize the economic burden on bidders. Here, efficient implies maximizing social welfare, truthful implies incentivizing truthful bids and collecting minimal payments from the bidders. 
\cite{TACCHETTI19} focuses on DL based approaches for auctions that incentivize truthful bids, maximize social welfare, and minimize the economic burden on buyers.
Theoretically, all these properties cannot be simultaneously satisfied. Hence, the authors design Neural Networks (NNs) to approximately achieve these properties. The NN architecture consists of Convolutional Neural Networks (CNNs) that learn Groves' payment rule for truthful mechanism. %MRB:Can explain why NNs are used here
The NN architecture is proposed for auctions with multiple units with decreasing marginal utilities. \cite{BRERO19} proposes a Machine Learning (ML) based Combinatorial Auction (CA) that improves preference elicitation based on value queries. In this method, each buyer's value function is learned via separate ML models (bundle-value pair as input) and these valuations are fed to another ML model that has the winner determination algorithm. The overall goal of the proposed model is to maximize the learned social welfare of all buyers. In \cite{WEISSTEINER20} the authors propose DL based iterative CAs where they first formulate a Deep Neural Network (DNN) based Winner Determination Problem as a mixed integer program problem and then compare the results with Support Vector Regression (SVR).

The authors in \cite{DENG22} propose obtaining reserve prices through ML techniques, which can be given as advice to participating buyers. They show that through this advice, buyers with large total values across auctions are limited in their power to manipulate the outcome of the auction. They also provide individual welfare bound on VCG auctions where reserve prices are approximated through ML techniques.

In the case of spectrum auctions, the main goal of the auctioneer is to maximize the total welfare of the participants, to create more jobs, to maximize connectivity, and to increase trade. 
%MRB:or maximize public welfare? (see sentence above)
\cite{TACCHETTI19} studies an efficient and truthful multi-unit auction Mechanism Design using NNs. It models multi-unit auctions as a VCG auction mechanism and proposes a model that uses the CNN architecture with loss function as the payment calculated using Groves' payment rule. The main objective of the paper is to incentivize truthful buyers and allocate goods that maximize the social welfare of all the buyers and minimize the economic burden on the buyers. This is achieved via designing truthful and efficient auctions that minimizes the sum of payments under the constraints of IR and weak budget balance and choosing the Groves payment rule by selecting a function $h:\textbf{V}_{-i} \rightarrow R$. The mathematical formulation is given below. 
For a set $M$ of $m$ items and a set $N$ of $n$ buyers, let $K$ be the set that consists of all possible ways to allocate the items to buyers. The utility of the buyer $i$ is $U_i(k,t_i) = v_i(k) - t_i$ where $v_i(k)$ is the valuation of the $k^{th}$ allocation by user $i$ and $t_i$ is the payment for user $i$.

\begin{equation}\label{eq:r2_obj}
    h^* = \argminA_{h \in \textit{H}} E_{v_i \sim \textbf{F}} \sum_{i=1}^n t_i
\end{equation}
such that

\begin{equation}\label{eq:r2_con1}
    \sum_{i=1}^n t_i \ge 0
\end{equation}

\begin{equation}\label{eq:r2_con2}
    U_i(k^*,t_i) \ge 0
\end{equation}

The valuation of each buyer is assumed to be sampled from an independent and identically distributed set $\textbf{F}$. The dataset of  $L$ realized $n$-player profiles $D = \{(v^l_1, v^l_2, ..., v^l_n): l= 1,2, \dots L\}$ is available. Eqs. \ref{eq:r2_con1} and \ref{eq:r2_con2} represent non-deficit and IR constraints which when combined with the objective function (Eq. \ref{eq:r2_obj}) formulate the Lagrangian function which is used as a loss function for the NN architecture shown in Eq. \ref{eq:r2_lagrangian}.

%VR-done: Not able to fix overfull box above
%Ans: Fixed with newline

\begin{dmath} \label{eq:r2_lagrangian}
    \hat{h} = \argminA_{h \in \textit{H}} \sum_{l=1}^L (\sum_{i=1}^n t^l_i + \lambda_b (\min\{\sum_{i=1}^n t^l_i ,0 \})^2 \\ + \lambda_r \sum_{i=1}^n ((\min\{U_i(k^*,t_i),0\})^2)))
\end{dmath}

Groves payment rule is selected by using the following two approaches.

\begin{enumerate}
    \item An NN is constructed to directly implement $\hat{h}$ given in equation \ref{eq:r2_lagrangian} and minimize this empirical loss given a data set of realized profiles $D$.
    \item An NN is constructed to learn a VCG redistribution function $r(v_1,\dots, v_{i-1},v_{i+1},\dots, v_n)$. Let $\hat{h}(.) = h_{VCG}(.) - r(.)$. Here, IR can be achieved by ensuring $r$ takes non-negative values and the VCG mechanism is already IR and the buyers' utilities will increase by giving back payments. 
\end{enumerate}

\begin{figure*}[h]
    \centering
    \includegraphics[scale=0.6]{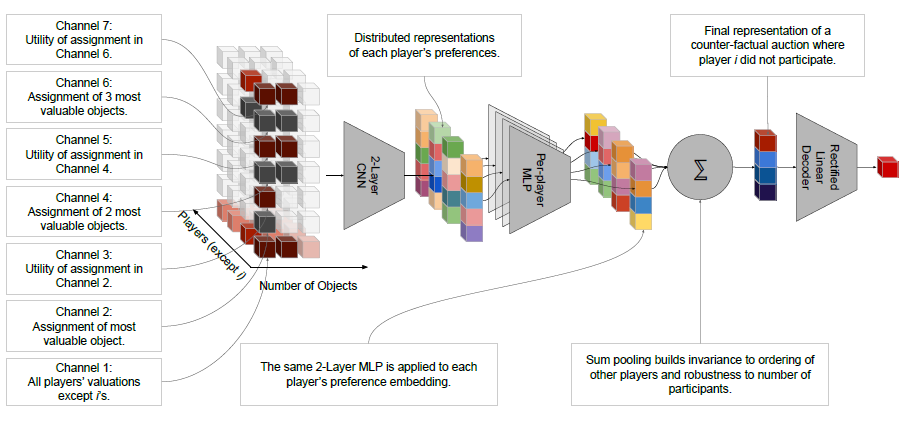}
    \caption{Multi-item auction with social welfare optimization using CNN (Figure reproduced from \cite{TACCHETTI19})}
    \label{fig:socialwelfare_1}
\end{figure*}

The same network architecture (Fig. \ref{fig:socialwelfare_1}) is used for the above settings. In computing the player $i$'s payment $t_i$ (based on Groves' payment rule), the network has to access reports from the other players but has no knowledge of the $i^{th}$ player's valuation. 
%the network architecture is designed to input only the bid amounts and valuation vectors of the competing buyers. 
By processing these competitor profiles, while ignoring the buyer's own bid, the network learns to approximate the Groves payment term. The proposed architecture is robust to change in the number of buyers, order invariant of buyers and convolution over buyers.

The architecture in Fig. \ref{fig:socialwelfare_1} represents an auction with five buyers and three units to be sold 
($n=5$ and $|K|=3$) 
that have decreasing marginal utilities. The input to the architecture is a tensor of size $4 \times 3 \times 7$ 
($(n-1) \times |K| \times (2|K|+1)$) 
and is processed with a 2-layer CNN to extract a distributed representation of preferences per buyer where the preference embedding is performed using a 2-layer Multi Layer Perceptron. In order to ensure ordering invariance of buyers and robustness to the number of buyers, the resulting embeddings are sum-pooled which results in a single positive number that represents either $\hat{h}$ or a redistribution function $r()$. 

CAs such as spectrum auctions and allocation of ad-slots in TV are used to allocate multiple items in a bundle to multiple buyers. The main challenge is that when the number of items grow, the bundle size also grows exponentially. This causes a more challenging preference elicitation. \cite{BRERO19} proposes an auction mechanism that uses a traditional ML technique, known as \textit{Machine Learning Powered Iterative Combinatorial Auction (MLCA)} with the goal of maximizing empirical efficiency in a realistic CA. In this architecture, an ML algorithm for each buyer is trained on value reports generalizable to the whole bundle space. In every round of an auction, the MLCA algorithm uses ML-powered query sub-routines that contain the learned valuations for each buyer and the winner determination steps. The mathematical formulation of the model is given below. 

A bundle is a subset of items ($\subseteq M$). Let $\textit{X} = \{0,1\}^m$  denote a set of bundles that contains indicator vectors. For each bundle $x \in \textit{X}$, let $v_i(x)$ denote the $i^{th}$ buyer's true value for obtaining $x$. Let $F$ be the set of all feasible allocations for a given allocation vector $a = (a_1,\dots, a_n) \in \textit{X}^n$ and payments vector $p = (p_1,\dots,p_n) \in R^n$, the utility function of the buyer $i$ is then defined as

\begin{equation}\label{eq:ML_TruthElicitation_utility}
    u_i(a,p) = v_i(a_i) - p_i
\end{equation}

For the given allocation $a$, the social welfare is defined as $V(a) = \sum_{i \in N} v_i(a)$ and the measure of efficiency of allocation is defined as $\Gamma(a) = \frac{V(a)}{V(a^*)}$, where \\ $a^* \in \argmaxA_{a\in F} V(a)$. The process of designing an Iterative CA is as follows. For each buyer $i$, let a bundle value pair report be $R_i = \{(x_{ik},\hat{v}_{ik})\}_{k \in L}$, where the $k^{th}$ bundle is $x_{ik}$ with corresponding value report $\hat{v}_{ik}$ and $L = \{1,\dots,l\}$. Let $R = (R_1,\dots,R_n)$ be a profile of pair reports. For example, if a bundle $x \in R_i$, then the buyer $i$ reported a value for this bundle $x$. Let $F_R = \{a \in \textit{F}: a_i \in R_i \; \forall i\}$ be the feasible report set and $\hat{v}_i()$ be the report function with $\hat{v}_i(x) = -\infty$ if $x \notin R_i$. For a given report set $R$, the mechanism computes the efficient allocation so as to maximize social welfare (i.e., $\argmaxA_{a \in F_R} \sum_{i \in N} \hat{v}_i$). From the Vickrey auction, every buyer has to report the value for every bundle which is captured in the report function $\hat{v}()$ and the outcome is computed using the following rules.

\begin{enumerate}
    \item \textbf{\textit{The allocation rule}} : $a^{VCG} \in \argmaxA_{a \in F} \sum_{i\in N} \hat{v}_i$
    \item The $i^{th}$ buyer \textbf{\textit{payment}} is: $p_i^{VCG} = \sum_{j \in N\setminus\{i\}} \hat{v}_j(a_j^{-i}) - \sum_{j \in N\setminus\{i\}} \hat{v}_j(a_j^{VCG})$, \\ where $a^{-i} \in \argmaxA_{a in F} \sum_{j \in N\setminus\{i\}} \hat{v}_j(a_j^{-i})$.
\end{enumerate}

\begin{figure}[h]
    \centering
    \includegraphics[scale=0.48]{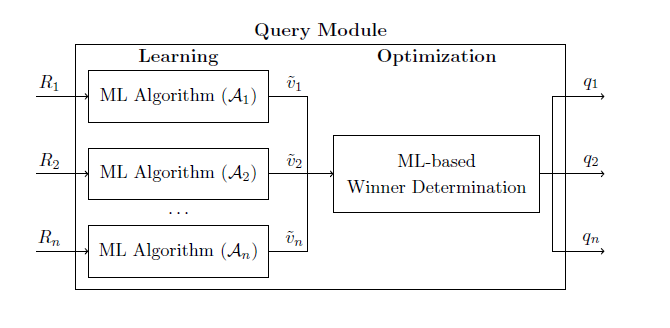}
    \caption{Schematic Representation of Query module (Figure reproduced from \cite{BRERO19})}
    \label{fig:MLCA}
\end{figure}

The schematic representation of MLCA is given in Fig. \ref{fig:MLCA}.

The main goal of MLCA is to provide a good incentive and to maximize empirical efficiency in real time CA. It consists of the query module given in Fig. \ref{fig:MLCA} as a sub-routine in every auction round. 
The MLCA can be thought of as an iterative VCG mechanism with a constraint on the maximum amount of information exchange between the buyer and the seller. 
The ML powered query module consists of the learning module that learns the valuation of each buyer and the winner determination module that computes the allocation vector. The input to the query module is set of bundle-value pairs $R_i$ for the $i^{th}$ buyer and for each buyer $i$ an ML algorithm $\textsc{A}_i$ is trained with these pairs to predict the valuation known as learned valuation. An ML based winner determination module which uses a kernel based SVR predicts the allocation by maximizing the learned social welfare formulated from predicted learned valuations of all buyers. This allocation is used as a query profile (which is a feasible allocation) for gathering bids from the buyers for the final allocation. 

The MLCA algorithm uses the query module sub-routine to compute feasible allocations and obtains the bids from the buyers through an iterative process. From the collected bids, it computes the final allocation and each buyers' payments using the equations given above in the outcome computation rules. As opposed to SVR for the winner determination problem in \cite{BRERO19}, the authors of \cite{WEISSTEINER20} use a DNN based winner determination problem for the Iterative CA that has the same problem formulation as that of \cite{BRERO19}. Let $\hat{v}^t_i$ be the true value estimator of an $i^{th}$ buyer and the data set to be trained be $B^{t-1}_i$ (i.e,. queried values are up to round $t-1)$. Then, the estimated social welfare function is

\begin{equation}
    \hat{V}^t = \sum_{i\in N} \hat{v}^t_i
\end{equation}

For every $i \in N$, the value estimate $\hat{v}^t_i$ is modelled using fully connected DNN as $\textit{N}_i:\{0,1\}^m \rightarrow R_{+}$ and the estimated social welfare function becomes

\begin{equation}
    \hat{V}^t = \sum_{i\in N} \textit{N}_i
\end{equation}

\begin{figure*}[h]
    \centering
    \includegraphics[scale=0.6]{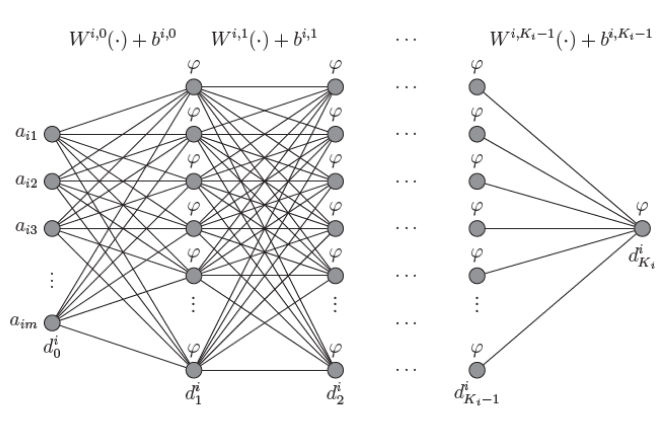}
    \caption{Representation of a DNN $\textit{N}_i$ (Figure reproduced from \cite{WEISSTEINER20})}
    \label{fig:DNN_v1}
\end{figure*}

The estimated value function is modelled using DNNs as shown in Fig. \ref{fig:DNN_v1}. The network takes $m$ item bundles of inputs $x \in \{0,1\}^m$ at each time instant $t$ and outputs a real number. Each DNN $\textit{N}_i$ has $K-i - 1$ hidden layers ($K_i \in N$) with $d^i_i$ hidden nodes in the $k^{th}$ hidden layer where $k \in \{1,\dots, K_i-1\}$. All hidden layers use a ReLu activation function and an the output layer uses a linear activation function since it is a regression problem. The weight parameters are estimated using an ADAM optimizer.

%%%%%%%%%%%%%%%%%%%%%%%%%%%%%%%%%%%%%%%%%%%%%%%%%%%%%%%%%%%
%%%%%%%%%%%%%%%  New Section 
%%%%%%%%%%%%%%%%%%%%%%%%%%%%%%%%%%%%%%%%%%%%%%%%%%%%%%%%
\section{Learning Mechanisms with Fairness in Allocation}\label{sec:fair}

\cite{KUO20} introduces a fairness condition in addition to revenue maximization and incentive compatibility to design an auction using a DL model (ProportionNet). This technique is an extension to RegretNet proposed in \cite{DUTTING23}. \cite{BARMAN19} studies \textit{Envy freeness up to one good (EF1), maximin share guarantee (MMS)} and \textit{Nash social welfare (NSW)} fairness conditions. Using these metrics, \cite{AHMAD21} proposes an NN model for an auction whose allocation is fair as well as IC and IR while maximizing the social welfare metric.

There are some scenarios wherein the auctioneer may be interested in the fair allocation of resources to buyers, for example, the allocation of resources such as spectrum allocation and the online advertisement of career opportunities to all the users irrespective of categories. 
\cite{KUO20} studies revenue maximizing auctions that have strategy proof and fairness constraints for online advertisement auction scenarios. It extends the work of RegretNet studied in \cite{DUTTING23}. First, let us define one of the fairness metrics known as total variation of fairness \cite{ILVENTO20}. 

Let $\textsc{U}$ be the set of users who arrive randomly to visit the website and let $\textsc{A}$ be the set of advertisers. Let us consider $p_{i,u}$ to be the probability of allocation of advertiser $i$ to user $u$ and let us divide set $\textsc{A}$ into $c$ classes as $C = \{C_1,\dots, C_c\}$. For any class $m$, $1 \le m \le c$, consider the user-pair distance metric $d^m:U\times U \rightarrow [0,1]$ . Then the \textbf{total variation fairness} is an extension of an individual fairness where similar users are treated similarly and is defined as follows.

\begin{definition} \label{fairness}
An auction mechanism satisfies \textbf{total variation fairness}, if the $L_1$ distance between two user allocations is less than the $L_1$ distance between the two users. That is

\begin{equation*}
\sum_{i \in C_m} |p_{i,u} - p_{i,u'}| \le d^m(u,u') \quad \forall m \in \{1,\dots, c\} \; \forall u,u' \in U
\end{equation*}

\end{definition}

From Definition \ref{fairness}, we define the \textbf{unfairness} metric which is used as a loss function for the NN. It is defined as the amount of violation of total variation fairness by an auction mechanism for each user $u\in U$ and is given as

\begin{dmath}
    Unf_u = \sum_{u' \in U} \sum_{C_m \in C} max(0,(\sum_{i \in C_m} max(0,p_{i,u} - p_{i,u'})-d^m(u,u')))
\end{dmath}

The loss function is defined as

\begin{equation}
    L_{unf} = \sum_{u \in U} \lambda_{f,u} Unf_u + \frac{\beta_f}{2}(\sum_{u \in U} Unf_u)^2
\end{equation}

The fairness constraint is ensured by adding the loss function $L_{unf}$ with the loss function defined in Eq. \ref{eqn:legrangianwithoutbcp} for the RegretNet model (Fig. \ref{fig:regretnet}) \cite{DUTTING23} using the augmented Lagrangian approach. 

In many real world applications we consider the fair allocation of resources, for example, in course assignments, ad-auctions, and inventory pricing. In \cite{BARMAN19}, the authors consider the auctioning of indivisible goods to multiple strategic buyers in a single-parameter environment. They study EF1, MMS, and NSW %\textit{Envy freeness up to one good (EF1), maximin share guarantee (MMS)} and \textit{Nash social welfare (NSW)} 
fairness conditions. Let there be $n$ buyers and a single seller having $m$ goods to sell. Let $v_i \in R_+$ be the valuation for buyer $i$ (private value) and let the known public value summarization function among all buyers be $W: 2^{[m]}\rightarrow R_+$. For a given subset $S \subseteq [m]$, the valuation of the $i^{th}$ buyer is $v_i(S) = v_i W(S)$. Then, the fairness conditions are defined below.

\begin{definition} [\textbf{EF1: Envy freeness upto one good}] \label{def:ef1}
    Under this comparative notion of fairness, an allocation is said to be \textbf{EF1} iff, every buyer $i$ values their allotted set of goods $S_i$ as much as any other buyer $j$ values their allotted set of goods $S_j$, up to the removal of the most valuable good from the other buyer set and for some $g \in S_j$. It is mathematically given as
    
    \begin{equation*}
        \max_{(S_1,\dots,S_n)} \sum_{i=1}^n v_i W(S_i)
    \end{equation*}
    such that
    
    \begin{equation*}
        W(S_i) \ge W(S_j) -W(g) \quad \forall i,j \in [m] 
    \end{equation*}
    Further, the allocation is both EF1 and Pareto efficient for additive valuations. 
\end{definition}

\begin{definition} [\textbf{MMS:Maximin share guarantee}]
    Under this fairness notion, an allocation is said to be \textbf{MMS}, iff every buyer gets a bundle of value $\ge \epsilon$, where $\epsilon$ is an agent specific fairness threshold. Mathematically, it is given as
    
    \begin{equation*}
        \max_{(S_1,\dots,S_n)} \sum_{i=1}^n v_i W(S_i)
    \end{equation*}
    such that 
    
    \begin{equation*}
        W(S_i) \ge \epsilon \quad \forall i \in [m] 
    \end{equation*}
    
\end{definition}

\begin{definition} [\textbf{NSW: Nash Social Welfare}]
    It is defined as the geometric mean of the buyers valuations for their bundles. Mathematically, it is given as 
    
    \begin{equation*}
         \max_{(S_1,\dots,S_n)} (\Pi_{i=1}^n v_i)^{\frac{1}{n}}(\Pi_{i=1}^n W(S_i))^{\frac{1}{n}}
    \end{equation*}
   and it quantifies the amount of fairness in an allocation. An allocation that maximizes NSW is 
   %MRB:Remove word 'both' for extra line
   both 
   fair and Pareto efficient.
\end{definition}
   
Motivated by the fairness metrics of \cite{BARMAN19}, \cite{AHMAD21} proposes an NN model for an auction that has a fair allocation in addition to being IC and IR while maximizing social welfare. The proposed model is described below.

Let there be $n$ buyers with $N = \{1,\dots,n\}$ and $m$ items to sell with $M = \{1,\dots,m\}$. Consider the valuation of buyer $i$ to be $v_i(S):S \rightarrow R \; \forall S \subseteq M$. We assume that this is an additive valuation. Let the allocation be $\textbf{p}^w(\textbf{v})$ and the payment be $\textbf{t}^w(\textbf{v})$. Then the regret can be defined from Eq. \ref{eq:regret} where $u^w_i = v^T_ip^w_i - t^w_i$. Similarly, the dissatisfaction metric (i.e,. amount of dissatisfaction for each buyer $i$) is defined as

\begin{equation}  \label{eq:dissatisfaction}
    D_i = \max_{j \in N} u^w_j - u^w_i
\end{equation}

The proposed formulations are described below.

\subsubsection{Formulation 1}
This formulation considers Maximizing social welfare under the constraints of zero regret, IR and dissatisfaction free allocation. Mathematically, it is given as

\begin{equation}
    \max_w \frac{1}{L}\sum_{l=1}^L \sum_{i=1}^n (v^l_i)^T p^w_i
\end{equation}
such that

\begin{equation}
    rgt_i(w) = 0 \quad \forall i\in N
\end{equation}

\begin{equation}
    D_i =0 \quad \forall i\in N
\end{equation}
The loss function to the NN optimizer is given by

\begin{dmath}
    L_1 = \frac{1}{L}\sum_{l=1}^L \sum_{i=1}^n (v^l_i)^T p^w_i + LagrangeLoss \\ + RegretPenalty + DissatisfactionPenalty
\end{dmath}

where
\begin{equation}
    LagrangeLoss = \sum_{i=1}^n \lambda^i_{rgt} \widehat{rgt}_i + \sum_{i=1}^n \lambda^i_D \widehat{D}_i
\end{equation}

\begin{equation}
    RegretPenalty = \rho_{rgt} \sum_{i=1}^n (\widehat{rgt}_i)^2
\end{equation}

\begin{equation}
    DissatisfactionPenalty = \rho_D \sum_{i=1}^n (\widehat{D}_i)^2
\end{equation}

\begin{equation}
    \widehat{rgt}_i = \frac{1}{L}\sum_{i=1}^L rgt^l_i
\end{equation}

\begin{equation}
    \widehat{D}_i = \frac{1}{L}\sum_{i=1}^L D^l_i
\end{equation}

\subsubsection{Formulation 2}
\begin{figure}[h]
    \centering
    \includegraphics[scale=0.6]{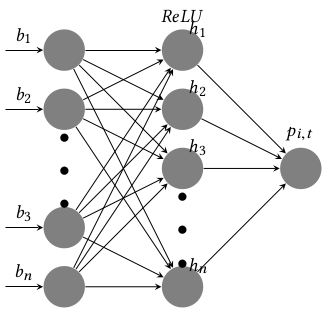}
    \caption{NN for Groves mechanism payment computation for each buyer $i$ (Figure reproduced from \cite{AHMAD21})}
    \label{fig:grovespayment}
\end{figure}

This formulation starts with the efficient allocation rule and tries to find a payment based on Groves' theorem. 
The idea here is to parameterize the function $H(v_{-i})$ of Grove's Mechanism using the Feed Forward Neural Network (FFNN) given in Fig. \ref{fig:grovespayment}.
%ANOTHER ALTERNATIVE:
%The idea here is to parameterize an alternative function to replace the Clarke tax using the Feed Forward Neural Network (FFNN) given in Fig. \ref{fig:grovespayment}.

%The idea here is to parameterize the function $H(v_{-i})$ given in Eq. \ref{eq:groves} using the Feed Forward Neural Network (FFNN) given in Fig. \ref{fig:grovespayment}.
%MRB: Groves equation referenced above, but commented out

Mathematically the formulation is given as

\begin{equation}
    \min_w \frac{1}{L} \sum_{l=1}^L (\sum_{i=1}^n D_i)^2
\end{equation}
such that

\begin{equation}
    I^i_r = 0 \quad \forall i \in N
\end{equation}

\begin{equation}
    I^i_{npt} =0 \quad \forall i \in N
\end{equation}

where the IR of user $i$ is $I^i_r = max\{0, t^w_i - v_i\}$ and No Positive Transfers is given by $I^i_{npt} = max\{0,t^w_i\}$. The loss function to the NN optimizer is given by

\begin{dmath}
    L_2 = \frac{1}{L} \sum_{l=1}^L (\sum_{i=1}^n D_i)^2 + LagrangeLoss \\ + IRPenalty + NPTPenalty
\end{dmath}

where
\begin{equation}
    LagrangeLoss = \sum_{i=1}^n \lambda^i_r (I^i_r) + \sum_{i=1}^n \lambda^i_{npt} I^i_{npt}
\end{equation}

\begin{equation}
    IRPenalty = \rho^i_r \sum_{i=1}^n (I^i_r)^2
\end{equation}
\begin{equation}
    NPTPenalty = \rho^i_{npt} \sum_{i=1}^n (I^i_{npt})^2
\end{equation}

\subsubsection{Formulation 3}
This formulation focuses on maximizing the NSW by finding Partitions as given below.

%MRB: Do we want to explain what are \Pi and S_i
\begin{equation*}
    \max_{(S_1,\dots,S_n)} (\Pi_{i=1}^n v_i)^{\frac{1}{n}}(\Pi_{i=1}^n w(S_i))^{\frac{1}{n}}
\end{equation*}

The partition is parameterized through an FFNN that is similar to the allocation module of RegretNet (Fig. \ref{fig:regretnet}). This module takes the public value summarization as input and outputs an $n \times m$ matrix that represents the partition of items. To ensure that the output matrix columns are almost one hot, the module uses the softmax function. The mathematical formulation is shown below.

\begin{equation}
    \max_W \frac{1}{L} \sum_{l=1}^L (\Pi_{i=1}^n v^l_ip_i)^{\frac{1}{n}}
\end{equation}
such that
\begin{equation}
     rgt_i(w) = 0 \quad \forall i\in N
\end{equation}

The corresponding loss function for the NN optimizer is given as 

\begin{dmath}
    L_3 = \frac{1}{L} \sum_{l=1}^L (\Pi_{i=1}^n v^l_ip_i)^{\frac{1}{n}} + LagrangeLoss + RegretPenalty 
\end{dmath}

where 

\begin{dmath}
    LagrangeLoss = \sum_{i=1}^n \lambda^i_{rgt} \widehat{rgt}_i
\end{dmath}

\begin{dmath}
    RegretPenalty = \rho_{rgt} \sum_{i=1}^n (\widehat{rgt}_i)^2
\end{dmath}

\begin{dmath}
    \widehat{rgt}_i = \frac{1}{L}\sum_{i=1}^L rgt^l_i
\end{dmath}

\cite{MISHRA22} introduces the concept of EEF1 Allocation (efficient and envy-free up to one item). The formulation is as follows. Consider a set $M = [m]$ of indivisible items to be allocated among $N = [n]$ buyers. Each agent $i \in N$ has a valuation function $v_i : 2^M \rightarrow R$. The valuation profile is represented as $(v_1,v_2,...,v_n)$ and the valuations are additive. The valuation of a buyer $i$ for a bundle $A_i$ is $v_i(A_i) = \sum_{j \in A_i} v_i(\{j\})$. The Utilitarian Social Welfare (USW) is defined as $sw(A,v) = \sum_{i \in N} v_i(A_i)$ and $v_i$ is assumed to be drawn from known distributions for each agent. The output allocation is $A \in \{0,1\}^{n\text{x}m}$ where $A_i \in [m]$ is the bundle assigned to buyer $i$. $A_i \bigcap A_k = \phi \; \forall i,k \in N$ and $\bigcup_i A_i = M$.

\begin{definition} [\textbf{Maximum Utilitarian Welfare (MUW)}] \label{def:muw}
An allocation $A^*$ is said to be MUW if it maximizes the USW. That is,
\begin{equation}
A^* \in \argmaxA_{A \in \{0,1\}^{n\text{x}m}} sw(A,v)
\end{equation}
\end{definition}

\begin{definition} [\textbf{EEF1 Allocation}] \label{def:eef1}
An allocation is said to be EEF1 if it satisfies EF1 fairness and is MUW amongst EF1 allocations (Def. \ref{def:ef1}).
\end{definition}

%The authors in \cite{MISHRA22} aim to achieve an allocation which they term EEF1 Allocation (efficient and envy-free up to one item) (Def. \ref{def:eef1}) using neural networks and train their network on allocation only, not payments. They formulate the problem as follows. Consider a set $M = [m]$ of indivisible items to be allocated among $N = [n]$ buyers. Each agent $i \in N$ has a valuation function $v_i : 2^M \rightarrow R$. The valuation profile is represented as $(v_1,v_2,...,v_n)$ and the valuations are additive. The valuation of a buyer $i$ for a bundle $A_i$ is $v_i(A_i) = \sum_{j \in A_i} v_i(\{j\})$. The Utilitarian Social Welfare (USW) is defined as $sw(A,v) = \sum_{i \in N} v_i(A_i)$ and $v_i$ is assumed to be drawn from known distributions for each agent. The output allocation is $A \in \{0,1\}^{n\text{x}m}$ where $A_i \in [m]$ is the bundle assigned to buyer $i$. $A_i \bigcap A_k = \phi \forall i,k \in N$ and $\bigcup_i A_i = M$. The NN architecture is shown in Fig. \ref{fig:eef1_arch}. 

\begin{figure}[h]
    \centering
    \includegraphics[scale=0.4]{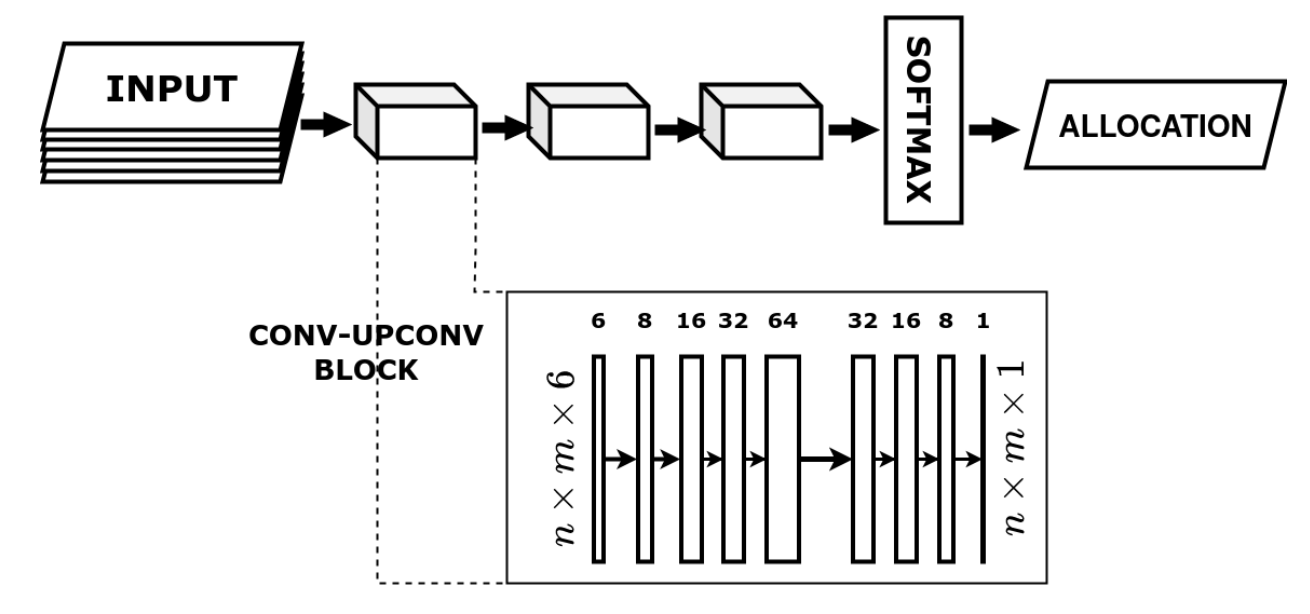}
    \caption{EEF1-NN Architecture (Figure reproduced from \cite{MISHRA22})}
    \label{fig:eef1_arch}
\end{figure}
%\begin{figure*}[h]
%    \centering
%    \includegraphics[scale=0.6]{eef1_arch.png}
%    \caption{EEF1-NN Architecture (Figure reproduced from \cite{MISHRA22})}
%    \label{fig:eef1_arch}
%\end{figure*}

%\begin{definition} [\textbf{EEF1 Allocation}] \label{def:eef1}
%An allocation is said to be EEF1 if it satisfies EF1 fairness and is MUW (Def. \ref{def:muw}) amongst EF1 allocations (Def. \ref{def:ef1}).
%\end{definition}

%\begin{definition} [\textbf{Maximum Utilitarian Welfare (MUW)}] \label{def:muw}
%An allocation $A^*$ is said to be MUW if it maximizes the USW, i.e.
%\begin{equation}
%A^* \in \argmaxA_{A \in \{0,1\}^{n\text{x}m}} sw(A,v)
%\end{equation}
%\end{definition}

The NN architecture used in \cite{MISHRA22} is shown in Fig. \ref{fig:eef1_arch}. The optimization problem is formulated as

\begin{equation} \label{eq:eef1_opt1}
    \text{Minimize } -E_v[sw(A,v)] = E_v[\sum_{i \in N} v_i(A_i)]
\end{equation}

\begin{equation} \label{eq:eef1_opt2}
    \text{subject to } E_v[\sum_{i \in N} ef1_i(A,v)] = 0
\end{equation}
where

\begin{dmath} \label{eq:eef1_op31}
    ef1_i(A,v) = \sum_{k \in N} \text{max}\{0,(v_i(A_k) - v_i(A_i)) \\ + \text{min}\{-\max_{j \in A_k} v_i(\{j\}, \min_{j \in A_i} v_i(\{j\}))\}\}
\end{dmath}

Consider there to be $L$ samples of valuation profiles $(v^1,v^2,...v^L)$, the loss for each sample is then given by
\begin{dmath} \label{eq:eef1_loss1}
    Loss(v^l,w) = \frac{1}{n\text{x}m} [-sw(A^w(v^l),v^l) \\ + \lambda \frac{\sum_{i \in N} ef1_i(A^w(v^l),v^l)}{n}]
\end{dmath}
and the following loss is minimized with respect to the NN parameters, $w$.
\begin{equation} \label{eq:eef1_loss2}
    L_{EEF1}(v^l,w) = \frac{1}{L} \sum_l Loss(v^l,w)
\end{equation}

%%%%%%%%%%%%%%%%%%%%%%%%%%%%%%%%%%%%%%%%%%%%%%
%%%%%%%%%%%%%%%%%%  New Section 
%%%%%%%%%%%%%%%%%%%%%%%%%%%%%%%%%%%%%%%%%%%%%%
\section{Learning Mechanisms with Budget Balance}\label{sec:BudgetBalanced}
Auctions are required in some situations to purely elicit true valuations from the participants, for example, to allocate public goods where no money needs to be collected. \cite{MANISHA18} designs a neural network, for such situations, to redistribute the payments made back to the participants such that DSIC is not compromised. It considers a setting where $p$ public goods need to be allocated to $n$ competing agents where each agent wants at most one of these goods. If the goods are homogeneous, agent $i$ values each good at $v_i = \theta_i$ whereas if they are heterogeneous, the valuations for each good for agent $i$ is given as $v_i = (\theta_{i1},\theta_{i2},...,\theta_{ip})$. Allocative efficiency %(Section \ref{sec:alloceff}) 
is ensured by calculating the allocations analytically as a function of the submitted bids.

The Redistribution Mechanism (RM) is determined by a rebate function $r()$. In general, the mechanism which redistributes a higher value or has a higher rebate while maintaining DSIC is considered the better mechanism. Two evaluation metrics are used to select a mechanism. The first one, \textit{Optimal in Expectations} (OE), compares the rebate functions based on maximum expected total redistribution where, given the prior distributions over agents' valuations, the total expected redistribution can be calculated. The second one, \textit{Optimal in Worst-case} (OW), finds the optimal in worst-case redistribution mechanism, based on the lowest redistribution index it guarantees.

\begin{figure}[h]
    \centering
    \includegraphics[scale=0.6]{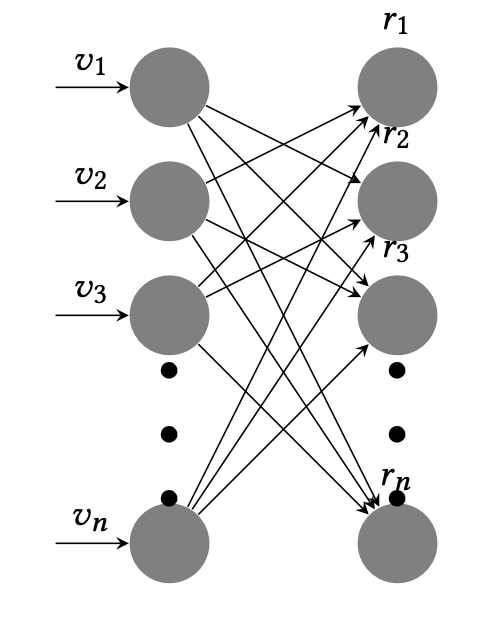}
    \caption{NN architecture for linear rebate function (Figure reproduced from \cite{MANISHA18})}
    \label{fig:linrebnn}
\end{figure}

\begin{figure}[h]
    \centering
    \includegraphics[scale=0.6]{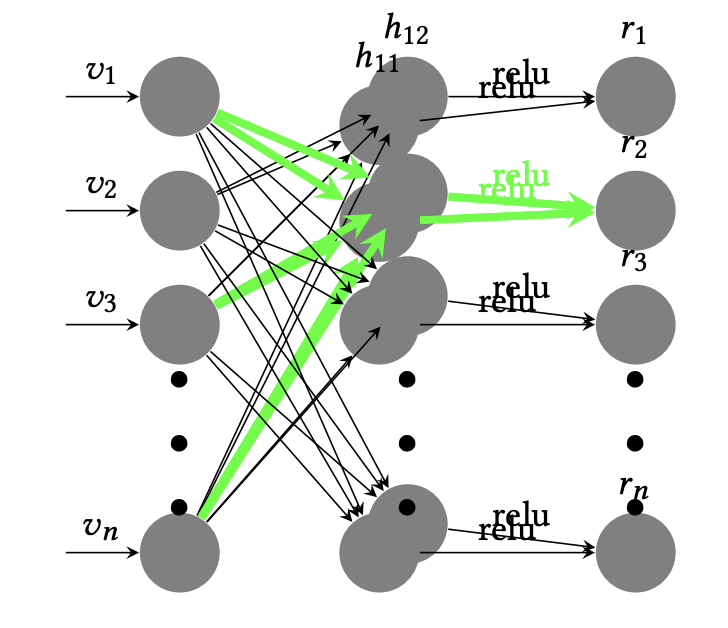}
    \caption{NN architecture for nonlinear rebate function (Figure reproduced from \cite{MANISHA18})}
    \label{fig:nonlinrebnn}
\end{figure}

The authors design two NN architectures, one with a linear rebate function and one with a nonlinear rebate function. The input nodes represent the valuation of the agents and the output nodes represent their rebates. The rebate for the $i^{th}$ agent is a function of the valuations of all the remaining agents. Considering there to be $n$ participating agents, there are $n$ input and output nodes and a total of $n-1$ weights and 1 bias since the weights and the bias used is the same for calculating the rebate for each agent. The linear rebate function and its NN architecure are shown in Eq. \ref{eq:linreb} and Fig. \ref{fig:linrebnn} respectively, and the nonlinear function and its NN architecture are shown in Eq. \ref{eq:nonlinreb} and Fig. \ref{fig:nonlinrebnn} respectively.

\begin{equation} \label{eq:linreb}
    r_i = \sum_{j=1}^{n-1} v_iw_j + b \; \forall i = 1,2,...,n
\end{equation}

\begin{equation} \label{eq:nonlinreb}
    r_i = \sum_{k=1}^h relu(\sum_{j=1}^{n-1} v_iw_{jk} + b)w_k^\prime + b^\prime \; \forall i = 1,2,...,n
\end{equation}
where $h$ is the number of hidden neurons and $relu(x) = max(0,x)$

The objective of the NN is to maximize the total rebate while obeying the feasibility and IR constraints. The feasibility constraint is given as

\begin{equation} \label{eq:feasibilityreb}
    g_1(w,b) : t - \sum_{i=1}^n r_i(w,b) \ge 0
\end{equation}
where $t$ is the total VCG payment by the agents and the IR constraint is given as

\begin{equation} \label{eq:irreb}
    %g^\prime (w,b) : r_i(w,b) \ge 0 \quad \forall i=1,2,...,n
    g_2(w,b) : r_i(w,b) \ge 0 \quad \forall i=1,2,...,n
\end{equation}
For the OE metric, the loss is defined as
\begin{equation} \label{eq:lossoereb}
    l(w,b) : \frac{1}{T} \sum_{j=1}^T \sum_{i=1}^n -r_i^j
\end{equation}
where $T$ is the total number of training samples. The feasibility constraint (Eq. \ref{eq:feasibilityreb}) can be modified as

\begin{equation} \label{eq:feasoereb}
    G^j(w,b) = max(-g(w,b),0) \quad \forall j = 1,2,...,T
\end{equation}
The overall loss function for the OE metric would then be
\begin{equation} \label{eq:totallossoereb}
    L(w,b) = l(w,b) + \frac{\rho}{2} \sum_{j=1}^T G^j(w,b)^2
\end{equation}
where $\rho$ is a constant. For the OW metric, given the worst possible redistribution index $k$, the loss is given by

\begin{equation} \label{eq:lossowreb}
    l(k) : -k
\end{equation}
such that
\begin{equation} \label{eq:g3owreb}
    g_3 : \sum_{i=1}^n r_i - kt \ge 0
\end{equation}

The feasibility constraint (Eq. \ref{eq:feasibilityreb}), IR constraint (Eq. \ref{eq:irreb}), and the inequality constraint for finding the worst case optimal (Eq. \ref{eq:g3owreb}) can be modified as shown in Eqs. \ref{eq:feasowreb},\ref{eq:irowreb},\ref{eq:worstowreb} respectively.

\begin{equation} \label{eq:feasowreb}
    G_1^j(w,b) = max(-g_1(w,b),0) \quad \forall j = 1,2,...,T
\end{equation}

\begin{equation} \label{eq:irowreb}
    G_2^j(w,b) = max(-g_2(w,b),0) \quad \forall j = 1,2,...,T
\end{equation}

\begin{equation} \label{eq:worstowreb}
    G_3^j(w,b) = max(-g_3(w,b),0) \quad \forall j = 1,2,...,T
\end{equation}
The overall loss function for the OW metric would then be

\begin{equation} \label{eq:totallossowreb}
    L(w,b,k) = l(k) + \frac{\rho}{2} \sum_{j=1}^T [G^j_1(w,b)^2 + G^j_2(w,b)^2 + G^j_3(w,b)^2]
\end{equation}

\section{Three Illustrative Applications}\label{sec:CaseStudies}

%%%%%%%%%%%  Uday - please include the paras that I have sent by email.
Deep learning as a technique to design complex mechanisms has been explored in many applications. Luong et. al. \cite{LUONG18} design an optimal auction in the context of edge computing resource management in mobile networks.  Luong and co-authors \cite{LUONG23} look into an interesting energy management application in vehicular networks using MyersonNet. Qian et. al. \cite{QIAN19} have applied the approach to resource allocation in wireless virtualisation. Bharadwaj and co-authors \cite{BHARDWAJ23} have explored an interesting input procurement application in agriculture for designing a versatile auction that satisfies incentive compatibility, individual rationality, cost minimization fairness of allocation, and certain business constraints. Zhou et. al. \cite{ZHOU18} have used the approach for predicting clickthrough rate in internet advertising while Liu et. al. \cite{LIU21} have explored the approach in e-commerce advertising. Golowich and co-authors \cite{GOLOWICH18} have used this approach for the multifacility location problem which arises in numerous contexts. Cepeda \cite{Cepeda} has investigated this approach in the context of electricity auctions where the author has extended the technique in Dutting et al. \cite{DUTTING23} to discover nearly optimal designs for electricity auctions incorporating capacity constraints and realistic assumptions.

We now discuss in some detail, the applications discussed by Luong et. al. \cite{LUONG18}, Qian et. al. \cite{QIAN19}, and Bhardwaj et. al. \cite{BHARDWAJ23}.

%There are several real-world instances where data-driven approaches to auction design have proven to be effective. Three such examples are explained below. %We consider two examples.
%The first is the problem of allocating resources in a mobile network and the second is procurement auctions in the agricultural setting.} \newline

\subsection{Efficient Energy Management in a UAV Assisted Vehicular Network}

\begin{figure*}[h]
    \centering
    \includegraphics[scale=0.6]{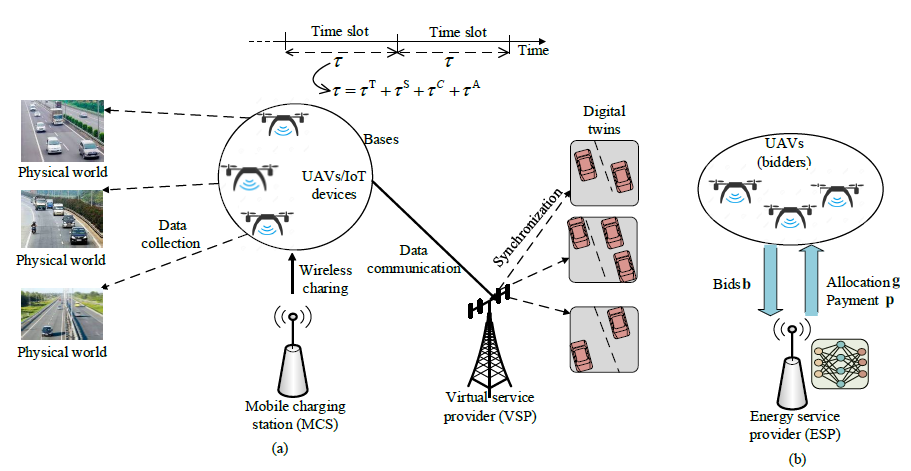}
    \caption{(a) A UAV assisted vehicular metaverse synchronisation system (b) A deep learning based auction model (Figure reproduced from\cite{LUONG23})}
    \label{fig:EMS_UAV}
\end{figure*}

%Digital twins play an important role in enhancing the quality of experience of users by creating digital aliases of physical entities. This is used to analyse the state of a physical entity continuously, monitor the health of an entity and provide recommendations. These digital entities are created in a Metaverse environment. 
In \cite{LUONG23}, the authors study efficient energy management using a neural network enabled auction mechanism. 
%The authors consider an Unmanned Aerial Vehicle (UAV) assisted framework for Metaverse creation of real physical entities as shown in  
In this example,  multiple unmanned aerial vehicles (UAVs) are deployed to collect data related to physical entities, such as vehicle information and roadway infrastructure. The collected data are then communicated; this %(data synchronisation is also considered) 
ensures data synchronization with a virtual service provider (VSP). The VSP creates, using this data, digital twins of the physical entities. Metaverse platform is used to create digital twins.  Fig. \ref{fig:EMS_UAV}.
As shown in the figure, each UAV returns to its base after gathering data, transmits this information to the VSP and charges itself via an Energy Service Provider (ESP). The ESP may contain one or more Mobile Charging Stations (MCSs). As the capacity of MCSs is limited and there is a revenue requirement for the ESP, the allocation of the UAVs must be done keeping these constraints in mind. 
One way of solving this energy management problem is to use traditional optimization methods with a set of constraints. However, this does not consider economic factors, such as profit and loss. The authors use a mechanism design approach by exploring auction based mechanisms that consider the profit, cost, and revenue.

The ESP is considered as a seller (or auctioneer) who provides MCS to UAVs and the UAVs are considered as buyers who bid to charge via a mobile charging station. The ESP  collects bids from UAVs and decides the winner and the winning price. Two scenarios are considered: the first is the ESP having a single mobile charging station and the second is the ESP having multiple MCS. For each data transfer activity, the UAV receives a reward, and it bids to charge from the ESP. 

Let $\lambda_i$ be the sensing rate (packets per second) of UAV $i$, let $D$ be the number of digital twins and let the decay rate be denoted by $\theta$. The reward function is then given by

\begin{equation}\label{eqn:rewardfn}
r_i  = \frac{D\eta(1+\theta)\lambda_i}{\sum_{k=1}^N \lambda_k}
\end{equation}

When $\eta = 0$, the VSP does not synchronise with all digital twins and their value will deteriorate at the rate of $\theta$. When $\eta>0$, the VSP  slows down the process of deterioration of digital twins. Further, the VSP will synchronise with all digital twins for a higher value of $\theta$. Greater incentives from the VSP to a UAV imply more rewards for the UAV. 
%When the incentives from the Virtual Service Provider to a UAV is more, it implies that the UAV's rewards increases. 
The following valuation function is considered. Let $E^T_i, E^S_i,$ and $E^C_i$ be the travelling energy, sensing energy, and communication energy of the $i^{th}$ UAV in a given time slot, respectively. The total energy then becomes

\begin{equation*}
    E^{Tot}_i = E^T_i + E^S_i + E^C_i
\end{equation*}
For a given scaling factor $\alpha \in (0,1)$, the valuation function for UAV $i$ is defined as

\begin{equation}
    v_i = \frac{(1 + r_i\frac{E^{Tot}_i}{E^R_i})^{1-\alpha}}{1-\alpha}
\end{equation}
where $E^R_i$ is the remaining energy in UAV $i$. %It is inversely proportional to the valuation and the -Commented out because E^R_i is not exactly inversely proportional to v_i, there is a power term in the equation and other variables as well.
The UAV with the lowest remaining energy will have the highest valuation and will bid accordingly. The two scenarios are described below.

\subsubsection*{Scenario 1: ESP with a single mobile charging station} 
In this case, the ESP with a single mobile charging station will act as a seller and multiple UAVs bid for the usage of the mobile charging station to charge. Myerson auction is used to obtain the optimal auction while ensuring Bayesian incentive compatibility and individual rationality.   may not guarantee optimal revenue to the ESP.  The proposed auction model uses MyersonNet, shown in Fig. \ref{fig:myersonnet}. The auction mechanism uses virtual bid prices which are computed as per Fig. \ref{fig:monotonefunctions}. The auction computes the allocation and payments as shown in Fig. \ref{fig:spa0}. 

\subsubsection*{Scenario 2: ESP with multiple mobile charging stations} 
\begin{figure}[h]
    \centering
    \includegraphics[scale=0.5]{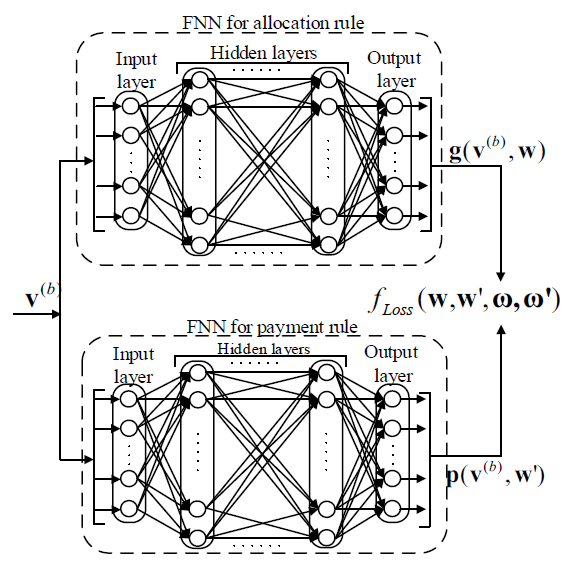}
    \caption{Feed forward neural network for allocation and payment rules (Figure reproduced from \cite{LUONG23})}
    \label{fig:UAV_FFNN}
\end{figure}

%MRB: Can remove () for space as it is covered in Scenario 1 already
Let there be $M$ energy units and $N$ UAVs. Let $\textbf{g}$ and $\textbf{p}$ be the allocation and payment vectors respectively. The revenue optimization problem for the ESP then becomes

\begin{equation}
    \max_{\textbf{p},\textbf{g}} Rev(\textbf{p})
\end{equation}
such that
\begin{equation}
    IC_i(\textbf{p},\textbf{g}) = 0 \quad \forall i \in \{1,...,n\}
\end{equation}
%MRB:N? or n?
\begin{equation}
    IR_i(\textbf{p},\textbf{g}) = 0 \quad \forall i \in \{1,...,n\}
\end{equation}
where $Rev(\textbf{p})$ is the expected revenue. The revenue optimization problem can be solved using the feed forward neural network architecture given in Fig. \ref{fig:UAV_FFNN}. 

\textcolor{black}{It is observed in \cite{LUONG23},  through simulations, that the proposed auction based scheme outperforms the classical second price auction in terms of revenue}.
%MRB:in terms of revenue, while adhering to the IC and IR constraints?

%%%%%%%%%% This section is put after \end{document}
%%%%%%%%%%%%%%%%%%
%\subsection{Resource Allocation in a Mobile Network}
\subsection{Resource Allocation in a Mobile Network}
\begin{figure}[h]
    \centering
    \includegraphics[scale=0.9]{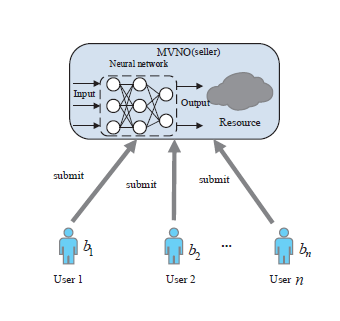}
    \caption{Auction for allocating sub-channels and power (Figure reproduced from \cite{QIAN19})}
    \label{fig:MVNO}
\end{figure}

A Mobile Network Virtual Operator (MNVO) provides services to its customers who do not own any wireless infrastructure. A wireless virtualization enables MNVO to use multiple virtual wireless networks to provide services over the same physical infrastructure. The allocation of resources owned by the MNVO to its mobile customers is a  challenging problem. The objective of MNVOs is to allocate resources (sub-channel and power) to maximize their respective revenues. Classical auction methods typically maximize social welfare and may often perform poorly in terms of revenues.  In \cite{QIAN19}, the authors present a deep learning network to learn an auction-based method to allocate resources  to the users of the MNVo. Fig. \ref{fig:MVNO} shows the proposed auction model with the MNVO auctioning the resources (sub-channel and power) to a set of $N = \{1,\dots,n\}$ users. Let user $i$ get power $p_i$ and sub-channel $sc_i$. 
If these values are higher than the corresponding values for other users $j \ne i$, the valuation of the $i^{th}$ user is given as $v_i = p_i \times sc_i$. Let the payment for user $i$ be $t_i$, then the utility is

\begin{equation} \label{eq:interimutiwithbc}
    u_i = 
\begin{cases}
    v_i - t_i,& \text{if buyer $i$ wins} \\
    0,              & \text{otherwise}
\end{cases}
\end{equation}

The authors propose a revenue optimal auction with incentive compatibility and individual rationality as the constraints. 
For this, input bids are first transformed to virtual valuations and then SPA-0 is applied to find allocation ($g^0$) and payment ($t^0$) rule. Let $\phi_i, i=1,\dots, n$ be the virtual transformation functions which are strictly monotone and let the MNVO revenue be $R(v,\phi,g^0,t^0) = t_i$, which is the payment from the winning buyer, $i$. Then the revenue maximization problem becomes the following.

\begin{equation}\label{eq:MVNO_uti}
    \max R(v,\phi,g^0,t^0)
\end{equation}
such that\\

\textbf{IR}: 

\begin{equation}\label{eq:MVNO_IR}
    u_i(v,\phi,g^0,t^0) \ge 0 \quad \forall i\in N
\end{equation}

\textbf{IC}:

\begin{equation} \label{eq:MVNO_IC}
    u_i((v_i,b_{-i}),\phi,g^0,t^0) \ge u_i((b_i,b_{-i}),\phi,g^0,t^0)
\end{equation}

The optimization problem of Eqs. \ref{eq:MVNO_uti}-\ref{eq:MVNO_IC} is solved using MyersonNet \cite{DUTTING23} given in Figs. \ref{fig:monotonefunctions},\ref{fig:myersonnet},\ref{fig:spa0}. \textcolor{black}{It is observed that the deep learning based resource allocation improves the revenue for MNVO as compared to Second Price Auction (with traditional optimization).}

%\newline
% related to another example
\subsection{A Procurement Auction for Agricultural Inputs}\label{subsec:AgriCaseStudy}
\cite{BHARDWAJ23} extends the RegretNet architecture \cite{DUTTING23} to tackle the problem of volume discount procurement auctions in the agricultural setting. Farmer Producer Organisations (FPOs) or Farmer Collectives (FCs) typically consist of around thousand farmers. Individual farmers may have requirements such as seeds, pesticides, fertilizers, etc., but may not be able to bargain the best deal from the suppliers if they approach the suppliers on their own. FPOs collect the requirements of all their farmers, organize all the requirements into clusters, and approach the suppliers with a bulk requirement. Due to the large volume of procurement, suppliers generally offer volume discounts, i.e., a higher percentage of discount on the original asking price as the number of units purchased increases. This is reflected in the bids submitted by the suppliers. 
In addition to the constraints satisfied by the RegretNet architecture, namely, IR, DSIC, and revenue maximization, the authors aim to satisfy certain additional constraints. 
%The authors aim to satisfy certain additional constraints in addition to IR, DSIC, and revenue maximization in the RegretNet architecture. 
Revenue maximization here becomes cost minimization, as it is a procurement auction. The additional constraints are fairness (envy-freeness), social welfare maximization, and business constraints (minimum and maximum number of winning suppliers, minimum amount of business to be given to each winning supplier, etc.). There is, however, some compromise to the social welfare maximization.

All these properties cannot be achieved simultaneously. Further, in the described agricultural setting, a large number of suppliers may participate in the auction, each selling a large number of units of each item. 
If an analytical approach such as linear programming were to be used, the computational complexity would significantly increase with an increase in the number of suppliers or units being sold. To overcome these challenges, a DL based approach is chosen. The architecture designed is an extension of the RegretNet architecture that is capable of handling volume discount bids and takes care of other constraints such as envy-freeness and business constraints.

The volume discount procurement auction setting of \cite{BHARDWAJ23} is as follows. Consider the case of a single buyer (FPO) who wishes to purchase $m$ homogeneous units of a single item from $n$ different suppliers. Let $m$ be divided into $k$ brackets. For each bracket, the bid submitted by the supplier (which is the price per unit in that bracket) must be less than or equal to the bid submitted by the supplier for the previous bracket. Each supplier $i$ submits a bid vector $b^{(i)} = (b_1^{(i)},b_2^{(i)},...,b_k^{(i)})$. Given all the input bids $b = (b^{(1)},b^{(2)},...,b^{(n)})$, the neural network outputs the allocation vector $a(b) = (a_1(b),a_2(b),...,a_n(b))$ and the payment vector $p(b) = (p_1(b),p_2(b),...,p_n(b))$ where $a_i(b)$ is the number of units sold by supplier $i$ and $p_i(b)$ is the payment made to supplier $i$.

The loss function is given as

\begin{dmath} \label{eq:agri1}
loss = cost + penalty_{regret} + penalty_{envy} \\ + penalty_{business} + LagrangianLoss
\end{dmath}

where

\begin{equation} \label{eq:agri2}
cost = \sum_{i=1}^n p_i(b),
\end{equation}
\begin{equation} \label{eq:agri3}
LagrangianLoss = \sum_{i=1}^n (\lambda_{regret}^{(i)} \tilde{r_i} + \lambda_{envy}^{(i)} e_i),
\end{equation}
\begin{equation} \label{eq:agri4}
penalty_{regret} = \rho_{regret} \sum_{i=1}^n \tilde{r_i}^2,
\end{equation}
\begin{equation} \label{eq:agri5}
penalty_{envy} = \rho_{envy} \sum_{i=1}^n e_i^2,
\end{equation}
\begin{equation} \label{eq:agri6}
penalty_{business} = 
\begin{cases}
    0, & \text{if $a^{(s)} \ge a_{min}$} \\
    \frac{\rho_{business}}{a^{(s)}}, & \text{if $a^{(s)}$ \textless $a_{min}$}
\end{cases}\\
\end{equation}

Through experimental analysis, the authors of \cite{BHARDWAJ23} show how a DL based procurement auction captures the essence of the agricultural market better than existing analytical methods. 
%The authors compare the results of their model to the popular VCG framework. 
Even as it minimizes additional constraints such as envy, when compared to VCG, their model saves the FPO more money.
%%%%%%%%%%%%%%%%%%%%%%%%%%%%%%%%%%%%%%%%%%%%%%%%%%
%%%%%%%%%%%%%%%%%%%%%%%%%%%%%%%%%%%%%%%%%%%%%%%%%%
%%%%%%%%%%%%%%%New Section %%%%%%%%%%%%%%%%%%%%%%
\section{Summary and Conclusions}\label{sec:Conclusion}
%MRB:Can use abbreviations in this section to save space

\begin{table*}[htb]
    \centering
    \begin{tabular}{|c|c|c|c|c|}
    \hline
    \textbf{Model} & \textbf{Strategyproofness} & \textbf{Efficiency} & \textbf{Fairness} & \textbf{Other Constraints} \\
    \hline
    \hline
    MenuNet & DSIC & Revenue & IR & - \\
    \hline
    % MoulineNet & & & & \\
    % \hline
    RochetNet & DSIC & Revenue & IR & - \\
    \hline
    RegretNet & DSIC & Revenue & IR & - \\
    \hline
    MyersonNet & DSIC & Revenue & IR & - \\
    \hline
    RegretFormer & DSIC & Revenue & IR & Regret Budget\\
    \hline
    Budgeted RegretNet & BIC and & Revenue & IR & Budget Constraint \\
     & Conditional IC & & & \\
    \hline
    % Deep Interest Network & & & & \\
    % \hline
    EquivarianceNet & DSIC & Revenue & IR & - \\
    \hline
    % PreferenceNet & & & & \\
    % \hline
    ProportionNet & DSIC & Revenue & IR, Envy- & - \\
     & & & freeness & \\
    \hline
    Redistribution Mechanism & DSIC & Social & IR & Budget Balance \\
     & &  Welfare & & \\
    \hline
    EEF1-NN & - & Social & IR, Envy- & - \\
     & & Welfare & freeness & \\
    \hline
    \end{tabular}
    \caption{Summary of properties satisfied by models reviewed in this paper}
    \label{tab:model_summary}
\end{table*}

In this paper, we have provided a state-of-the-art review of a recent approach, based on deep learning, for designing mechanisms that are theoretically impossible or difficult to achieve. 
We have provided the motivation for using a deep learning approach to auction design followed by a description of several key approaches that have emerged. The approaches presented are powerful and applicable for many real-world applications. We have illustrated the approaches using three applications: (a) efficient energy management in a vehicular network; (b) resource allocation in a mobile network; and (b) designing a volume discount procurement auction for agricultural inputs.

%MRB:This paragraph is all about \cite{ZHANG21}. It gives the impression that the current paper is just feeding off \cite{ZHANG21}.
%We could consider adding a paragraph before this paragraph, containing an overview of other works.
Table \ref{tab:model_summary} provides a  summary of the more prominent DL architectures for auction design as well as their evaluation. % \cite{ZHANG21}. 
The DL approaches, in current literature, used for designing auctions include MenuNet, Deep Interest Network, RochetNet, MoulineNet, RegretNet, PreferenceNet, Deep Neural Auction and the Reinforcement Learning setups. 
%MRB: List not matching Table II
%MRB:Dangling sentence below...commented
%The authors also discuss methods to tackle constraints and concerns that occur in industrial settings. 
MenuNet \cite{SHEN18} is based on the principle of taxation, which lets buyers make selections that yield an IC mechanism. In online auctions, the essential task of click through rate prediction is implemented using Deep Interest Network \cite{ZHOU18}. MoulineNet is used to determine agents' preferred optimal facility location \cite{GOLOWICH18}. An extension of RegretNet is PreferenceNet \cite{PERI21} that includes buyers' preferences in Auction design. The Reinforcement learning setup is based on a no-regret learning mechanism and in the Deep Neural Auction model \cite{LIU21}, the positional effect of ordering of sellers based on some metric is removed. The summary of key approaches referenced in our paper are given in Table \ref{tab:model_summary}. 
%Repeat of 1st sentence of same para. Should we remove the first one and add Zhang's reference here?
In this table, strategyproofness indicates the type of IC constraint the model aims to satisfy, efficiency indicates whether it is revenue maximizing or welfare maximizing, fairness indicates the model's aim to minimize violations of IR, envy, etc. and other constraints refer to constraints such as budget balance that the model aims to satisfy.

The topic presented in this paper provides a fertile ground for further research and innovation. The future research directions are theoretical as well as application driven.

On the theoretical front, an important topic for further investigation is interpretability and explainability of the deep learning based approach. There are not many results on this topic and this topic offers a technically challenging opportunity to advance this area. Another topic for further investigation is to consider different notions of fairness and different types of business constraints and investigate the design of mechanisms with these new notions. The third topic is regarding different auction types. %There is a multitude of auction types that have been employed. For example, combinatorial auctions are extremely popular. 
Extending the deep learning approach to other types of auctions like combinatorial auctions presents interesting challenges. As the fourth topic for future research, we wish to mention investigating the computational complexity of the deep learning based approach. A related topic will be to investigate computational platform and systems issues in solving the resulting deep learning problems.

On the application side, any practical application that currently employs a solution based on auctions is a potential candidate for applying the deep learning approach. Each application is unique in its own way and the desiderata of desirable properties will be different for different applications. Defining appropriate loss functions, coming up with the best neural network architecture for the given loss functions, and associated hyper parameter optimization will pose interesting research questions for each application. Can the design of a deep learning architecture for a given application be automated? That would be a grand challenge problem.

%MRB: To save space, throughout the paper, we can replace 'the authors in [] study' by '[] studies'. Similarly, 'In [], the authors' can be replaced by 'In []'
%MRB: Remove extra lines before and after every equation, table and image - this will save a lot of space!!

\section*{Declaration of Competing Interest}
The authors confirm that there are no financial conflicts of interest or personal relationships that could influence the reported findings in this manuscript.
\section*{Ethics approval and consent to participate}
Not applicable, as no ethical approvals or consent required for this article
\section*{Consent for publication}
Not applicable, as no individual patient data or identifying information is included in this manuscript.
\section*{Availability of data and material}
Not applicable, as no data is required for this study
\section*{Competing interests}
The authors have disclosed no competing interests related to this work.
\section*{Funding}
The first author V. Udaya sankar received funding from TARE (Teachers Associateship for Research Excellence) (File No: TAR/2021/000114) Grant of the Department of Science and Technology, Government of India.
\section*{Authors' contributions}
V. Udaya Sankar (VUS) initiated the study and drafted the initial manuscript. Vishisht, Mayank, Narahari all are responsible for providing crucial inputs and contributing to the major revision along with VUS.

%Ethics approval and consent to participate
%Consent for publication
%Availability of data and material
%Competing interests
%Funding       
%Authors' contributions
%Acknowledgements

\section*{Acknowledgements}
The authors would like to thank Prof. Sujit Gujar, Dr. Manisha Padala, and Mr. Ansh Das for offering detailed comments on this manuscript.

\bibliographystyle{IEEEtran}
\bibliography{authors}

\end{document}